\documentclass[preprint,12pt]{elsarticle}

\usepackage{amssymb}
\usepackage{amsmath}

\usepackage{lmodern}
\usepackage{tikz}
 
\usepackage{amsfonts}
\usepackage{titlesec}
\usepackage{dirtytalk}
\usepackage{braket}
\usepackage{amssymb}
\usepackage{braket}
\usepackage[caption = false]{subfig}
\usepackage{graphicx}
\usepackage{float}
\usepackage{hyperref}
\usepackage{acronym}
\usepackage{indentfirst}
\journal{Annals of Physics}

\begin{document}

\begin{frontmatter}



\title{\textbf{System Energies for Pentaquark Family Using Thomas-Fermi Quark Model}}


\author[a,b,e]{Suman Baral} 
\author[e,f]{Bipin Aryal}
\author[d,e]{Mohan Giri} 
\author[a]{Artem Denisenko} 
\author[c,e]{Gopi Chandra Kaphle}
\author[d,e]{Walter Wilcox}
\affiliation[a]{Division of Business and STEM, SUNY Niagara, Sanborn, NY,14132}
\affiliation[b]{Neural Innovations LLC, Lorena, TX 76655}
\affiliation[c]{Central Department of Physics, Tribhuvan University, Kirtipur}
\affiliation[d]{Department of Physics and Astronomy, Baylor University, Waco, TX 76798-7316}
\affiliation[e]{Everest Institute of Science and Technology, Kathmandu}
\affiliation[f]{Department of Physics & Astronomy, Louisiana State University, Baton Rouge, LA, 70803}
    \begin{abstract}
     
        This study calculates the system energies of families of pentaquarks, $uudc\overline{c}$, starting from 5 quarks up to 70 using Thomas-Fermi statistical quark model. The model assumes spherical symmetry with the particles continuously distributed with varying densities and boundaries. The particles interact with Coulombic forces and a discontinuity in the light quark density function is energetically favored. Total energies of the system are calculated for each multiquark system and compared to determine family stability characteristics.  
       A remarkable stability for multiquark combinations with multiples of four pentaquarks, which we term an icosaquark, is identified. The first icosaquark system is found to have lowest energy per quark, suggesting stability of a system consisting of four charm, four anticharm and twelve light quarks.  
    \end{abstract}



\begin{keyword}
Thomas-Fermi model, pentaquark, hadronic physics 


\end{keyword}

\end{frontmatter}



\section{Introduction}
    In the early 21st century, numerous efforts were made to experimentally verify tetraquarks~\cite{choi2003observation, hosaka2016exotic} and pentaquarks~\cite{aaij2015observation, lhcb2019observation}, and various models were proposed to study these exotic particles and their properties. In 2015, the LHCb Collaboration reported the discovery of two hidden-charm pentaquark states, $P_{c}(4380)$ and $P_{c}(4450)$, in the $J/\psi \, p$ invariant mass spectrum of $\Lambda^{0}_{b} \longrightarrow J / \psi K^{-} p$~\cite{aaij2015observation}. 
    Four years later, the LHCb Collaboration updated their results, proposing a new state called $P_{c}(4312)$, and splitting the previously observed $P_{c}(4450)$ into $P_{c}(4440)$ and $P_{c}(4457)$ states~\cite{lhcb2019observation}. 
    
    The discovery of these ($P_{c}^{+}$) pentaquarks has generated significant interest in the theoretical investigation of pentaquarks. Various explanations have been suggested, including the chiral soliton model (CSM)~\cite{scoccola2015pentaquark}, the diquark model~\cite{maiani2015new, lebed2015pentaquark, anisovich2015pentaquarks, guo2015reveal}, the QCD sum rule approach~\cite{chen2015towards, wang2016analysis}, the MIT bag model~\cite{madsen2003strangelets, madsen1993curvature}, and the Thomas-Fermi (TF) quark model~\cite{wilcox2009thomas, Liu:2012my}. Pentaquark existence results from lattice QCD are actually mixed\cite{Liu:2005yc}. While QCD serves as the fundamental theory of strong interaction, studying the structure of many-quark hadrons directly with lattice QCD is numerically challenging due to the non-perturbative nature of the theory~\cite{yan2022fully}.
    The TF quark model overcomes these limitations and provides insight into regions where lattice QCD calculations are inaccessible. It is the only theoretical tool available which can investigate {\it families} of multi-quark bound states such as pentaquarks. It is built on the TF version of density functional theory, and has been successfully applied to both atomic and nuclear physics.
    
    TF quark model with heavy and light particles produce heavy particles in the central region and light particles spread outside. In the case of the $P_C^+$ pentaquark, the $u$ and $d$ quarks are light particles that can have larger radii while the $c$ and $\bar{c}$ are the heavy quarks relative to the $u$ and $d$ quarks. This paper follows the research carried out by Giri {\it et al.}~\cite{giri2021investigation} by looking at the energy trend of family stability of pentaquark systems. 

\section{Theoretical Setting of the system}
    The Thomas-Fermi Quark Model is a semiclassical method which deals with spherical systems in which discrete particles are assumed to be continuously distributed with varying densities. This continuum of particles, made up of different colors and flavors, interacts via Coulombic forces. The dynamics of the particle distribution depend on distance, charge distribution, and the mass of the particles. As the system settles down, heavier quarks tend to spread around the center and the light ones get distributed throughout with a larger radius. The system is assumed to be at zero temperature, allowing us to disregard thermal energy.
    
     The model creates three different concentric spheres of varying densities and radius, two of which represent particles and one represents the antiparticle. This work is an extension from the previous work of our group~\cite{giri2021investigation}, where we studied the structure of same pentaquark system  $\overline{c}cuud$, and understood that characterizing such a pentaquark system actually requires three spheres. The three spheres represent light quarks (up \& , down), heavy quark (charm), and heavy antiquark (anti-charm), as charm and anticharm do not share the same boundary. We use three different functions for three different particles.
     
     We define the parameter $\eta$ to distinguish between the families of pentaquark system, where $\eta = 1$ corresponds to a single pentaquark system, $\eta = 2$ to a double pentaquark system (decaquark), $\eta = 3$ to a triple pentaquark system (pentadecaquark), and so on. We coined a name \say{pocket} for five particles. So, a decaquark has two pockets, a pentadecaquark has three pockets for example. One thing to note here is that we are not investigating the system energies of molecules but the system as a whole. Also, we understood in the previous paper that for $\eta = 1$ anti-charm has the smallest radius whereas for $\eta > 1$ charm has the smallest radius.
    
     In section \eqref{subsection:tfqm}, we apply the TF quark model to describe our system. This results in three equations, for three different TF functions. Then, in section \eqref{subsection:cons_condn} we assume that functions are linearly related if they share the same region and come up with consistency conditions that must be satisfied by these three functions. 
    Once these constants are determined, we calculate the kinetic, potential and vacuum energies contributions in section \eqref{subsection:total_energy} and derive the total energy of the system. This approach allows us to determine the energy stability trend for families of pentaquark systems.
      
    \subsection{Thomas-Fermi Pentaquark Model} \label{subsection:tfqm}
        By incorporating various color interactions among quarks and anti-quarks, along with their respective interaction probabilities, the TF quark model proves to be a valuable tool for describing the quark distribution within the pentaquark system. This is achieved through the utilization of Eqs. \eqref{eqn:tf_eqn_antiquark} and \eqref{eqn:tf_eqn_quark} from~\cite{giri2021investigation}.
        \begin{equation}
            \overline{\alpha}_I \frac{d^2\overline{f}_I(x)}{dx^2}= -\frac{6\eta}{5(5\eta-1)} \sum_J g_J \frac{(f_J(x))^\frac{3}{2}}{\sqrt{x}},
            \label{eqn:tf_eqn_antiquark}
        \end{equation} 
        \begin{eqnarray}
            \alpha_I \frac{d^2 f_I(x)}{dx^2} &=& -\frac{6\eta}{5(5\eta-1)}\sum_I \overline{g}_I \frac{\left( \overline{f}_I(x) \right)^\frac{3}{2}}{\sqrt{x}}
            - \frac{9\eta}{10(5\eta-1)}\left[\frac{(N_I g_I-1)}{N_I} \right. \nonumber \\
            &\times&\frac{\left( f_I(x) \right)^\frac{3}{2}}{\sqrt{x}} 
             + \left. \sum_{I \ne J} g_J \frac{\left( f_J(x) \right)^\frac{3}{2}}{\sqrt{x}} \right].
            \label{eqn:tf_eqn_quark}
        \end{eqnarray}
        Here, the indices $I$ and $J$ represent the flavor, with 1 corresponding to the light quarks (up and down) and 2 to the heavy quarks (charm and anti-charm). In the context of the TF density function, denoted as $f_I(x)$, a distinct function is assigned for each respective flavor. To comprehensively investigate the family stability of pentaquarks, three crucial TF functions are indispensable: $f_1(x)$ for light quarks (up and down), $f_2(x)$ for the heavy quark $c$, and $\overline{f}_{2}(x)$ for the heavy anti-quark $\overline{c}$. The parameters $\alpha_I$ and $\overline{\alpha}_I$ represent the ratios of light quark to heavy quark and light quark to heavy anti-quark masses, respectively. Additionally, $\eta$ denotes the number of pockets of pentaquarks, $g_J$ signifies the degeneracy of the flavor in a specific pocket $\eta$, and $N_I$ signifies the number of quarks with flavor $I$. Note that $x$ is the dimensionless radius and the physical radius, $r$, is given by the relation  $r = Rx$, where  
        \begin{equation}
            R=\frac{3 a}{8\alpha_{s}}  \left(\frac{ 3 \pi \eta}{2} \right)^\frac{2}{3},
            \label{eqn:radius_constant}
        \end{equation}
        and $ a=\hbar / m_{1} c$, $m_{1}$ is the mass of the light quark, and $\alpha_{s}$ is the strong coupling constant.
        
        Assuming different boundaries of heavy quarks and anti-quarks, a linear relationship between any two TF functions within each region is employed to eliminate the multiple unknowns encountered in the study's equations. Consistency conditions derived from Eqs. \eqref{eqn:tf_eqn_antiquark} and \eqref{eqn:tf_eqn_quark} are solved to obtain numerical solutions for these differentials in all regions. The solved proportionality constants associated with the functions $f_2(x)$ and $\overline{f}_2(x)$ across various pentaquark pockets followed two different trends. For $\eta$ = 1, the proportionality constant is found to be greater than one while for rest of the $\eta$ values, the proportionality constant is found to be less than one. This deviation from one results in a conclusion: charm and anti-charm do not have the same boundary. To adequately describe the distribution of these quarks, three different radii are essential for the three TF functions, facilitating a proper delineation of the entire distribution into three distinct regions~\cite{giri2021investigation}. 
        
        To account for the different trends of the distribution across different pentaquark pockets in~\cite{giri2021investigation}, our investigation is divided into two cases: \textbf{Case I} and \textbf{Case II}. In Case I, where we examine the pentaquark pocket with $\eta = 1$, the entire distribution is segmented into three distinct regions. These regions, shown in Figure \ref{fig:quark_distribution_three_region}(a), are defined as follows: The inner region spans $0 < x < \overline{x}_{2}$. It has all three TF functions non-zero. The middle region encompasses $\overline{x}_{2} < x < x_2$ which contains only $f_2(x)$ and $f_1(x)$ as $\overline{f}_2(x)$ becomes zero. The outermost region extends from $x_2 < x < x_1$ with only $f_1(x)$ remaining non-zero. 
        
        In Case II, where we investigate pentaquark pockets with $\eta = 2$ and values greater than 2, a similar division of the overall distribution, shown in Figure \ref{fig:quark_distribution_three_region}(b), is employed. However, there is a distinction in the definition of the regions: The inner region now covers $0 < x < x_2$ and has all the TF functions non-zero. The middle region occurs within $x_2 < x < \overline{x}_{2}$. On this region, unlike in Case I, $f_2(x)$ becomes zero. The outermost region remains defined as $\overline{x}_{2} < x < x_1$ with only $f_1(x)$ as non-zero. Here, $x_2$, $\overline{x}_{2}$, and $x_1$ denote the boundaries corresponding to the charm, anticharm, and light quarks (up and down), respectively. This segmentation allows for a detailed analysis of the quark distribution in different regions within the pentaquark system.  
        
        A linear relationship between any two TF functions within each region is employed to eliminate the multiple unknowns encountered in the study's equations. Consistency conditions derived from Eqs.~\eqref{eqn:tf_eqn_antiquark} and \eqref{eqn:tf_eqn_quark} are solved to obtain numerical solutions for these differentials in all regions for both Case I and Case II. This process leads to the determination of the TF density functions and the boundaries $x_2$, $\overline{x}_{2}$, and $x_1$. 
        
        The obtained TF density functions and boundaries are then utilized to analyze energy trends, physical radius, and other essential properties of pentaquarks across pockets $\eta$ ranging from 1 to 10.
        For the anti-charm, charm and light quarks, Eqs.~\eqref{eqn:tf_eqn_antiquark} and \eqref{eqn:tf_eqn_quark}, respectively, can be written in the extended form as Eqs.~\eqref{eqn:tf_eqn_anticharm}, \eqref{eqn:tf_eqn_charm} and \eqref{eqn:tf_eqn_light_quark}. These are used to obtain the consistency conditions in individual cases.
        \begin{eqnarray}
             \overline{\alpha}_2 \frac{d^2\overline{f}_{2}(x)}{dx^2}= -\frac{6\eta}{5(5\eta-1)} \left[ g_1 \frac{(f_1(x))^\frac{3}{2}}{\sqrt{x}}
             +  g_2 \frac{(f_2(x))^\frac{3}{2}}{\sqrt{x}} \right],
             \label{eqn:tf_eqn_anticharm}
        \end{eqnarray}
        
        \begin{eqnarray}
            \alpha_2 \frac{d^2 f_2(x)}{dx^2} &=&-\frac{6\eta}{5(5\eta-1)}    \overline{g}_{2} \frac{\left( \overline{f}_{2}(x) \right)^\frac{3}{2}}{\sqrt{x}} 
            - \frac{9\eta}{10(5\eta-1)}\left[\frac{(N_2 g_2-1)}{N_2} \right. \nonumber \\ 
            &\times& \frac{\left( f_2(x) \right)^\frac{3}{2}}{\sqrt{x}}
            + \left. g_1 \frac{\left( f_1(x) \right)^\frac{3}{2}}{\sqrt{x}} \right],
            \label{eqn:tf_eqn_charm}
        \end{eqnarray}
        
        \begin{eqnarray}
            \alpha_1 \frac{d^2 f_1(x)}{dx^2} &=& -\frac{6 \eta}{5(5 \eta -1)}\overline{g}_{2} \frac{(\overline{f}_{2}(x))^\frac{3}{2}}{\sqrt{x}}
            -\frac{9\eta}{10(5\eta-1)}\left[\frac{(N_1 g_1-1)}{N_1} \right. \nonumber \\ 
            &\times& \frac{\left( f_1(x) \right)^\frac{3}{2}}{\sqrt{x}}
            + \left. g_2 \frac{\left( f_2(x) \right)^\frac{3}{2}}{\sqrt{x}} \right].
            \label{eqn:tf_eqn_light_quark}
        \end{eqnarray}
        
    \subsection{Consistency conditions}
    \label{subsection:cons_condn}
        \begin{figure*}
            \centering
            \includegraphics[width = \textwidth]{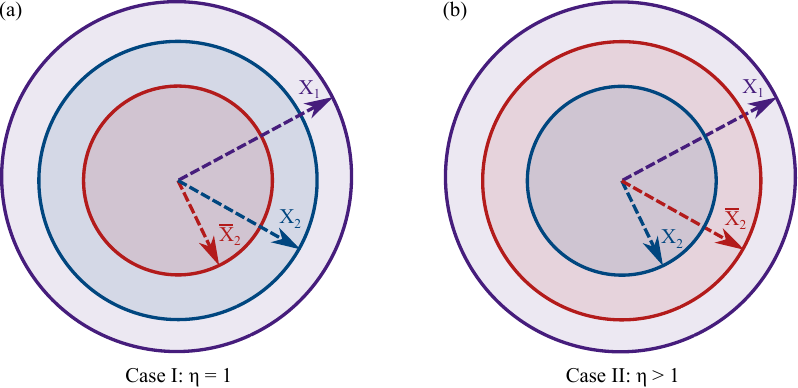}
            \caption{The schematic diagram of the distribution of quarks for the case of (a) $\eta = 1$, and (b) $\eta = 2-10$ in three different regions. $\overline{x}_{2}$, $x_{2}$ and $x_{1}$ differentiate the boundary between these regions.}
            \label{fig:quark_distribution_three_region}
        \end{figure*}
        In this section, we derive the consistency conditions required for our model to produce a single collective density of states. Following this, we outline the approach for selecting parameters to numerically solve these consistency conditions.
        
        Figure~\ref{fig:quark_distribution_three_region} illustrates the distribution of quarks across three distinct regions. Figure~\ref{fig:quark_distribution_three_region}(a) depicts the quark distribution for a system with 5 quarks and an $\eta$ value of 1. In Case I the anti-charm quark ($\overline{c}$) occupies the innermost region $0 < x < \overline{x}_{2}$, the charm quark ($c$) spans $0 < x < x_2$, and the light quarks ($u, d$) extend up to $x_1$ within $0 < x < x_1$.
        For systems with more than 5 quarks, as shown in Figure~\ref {fig:quark_distribution_three_region}(b), the distribution changes. Here, the charm quark ($c$) resides in the innermost region $0 < x < x_2$, the anti-charm quarks ($\overline{c}$) are in the innermost and central regions, and the light quarks ($u, d$) extend up to $x_1$. This is the scenario that we introduced as Case II.
        
        \subsubsection{Case I: $\eta =1$}        
            In the region $0 < x < \overline{x}_{2}$, we define three functions: $f_2(x)$, $\overline{f}_{2}(x)$, and $f_1(x)$. These functions are assumed to be linearly related as follows:  
            \[
            f_1(x) = \beta_{1\overline{2}} \overline{f}_{2}(x), \quad f_2(x) = \beta_{2\overline{2}} \overline{f}_{2}(x),
            \]
            where $\beta_{1\overline{2}}$ and $\beta_{2\overline{2}}$ are linear proportionality constants. 
            Given this relationship and the condition $\alpha_1 = m_{1}/m_{1} = 1$, Eqs.~\eqref{eqn:tf_eqn_anticharm}, \eqref{eqn:tf_eqn_charm}, and \eqref{eqn:tf_eqn_light_quark} simplify to Eqs.~\eqref{eq:f2_bar_first}, \eqref{eq:f2_bar_second}, and \eqref{eq:f2_bar_third}, respectively.
            \begin{eqnarray}
                   \frac{d^2\overline{f}_{2}(x)}{dx^2} &=& -\frac{6\eta}{5(5\eta-1) \overline{\alpha}_2} \biggl[ g_1 (\beta_{1\overline{2}})^\frac{3}{2}
                   + g_2 (\beta_{2\overline{2}})^\frac{3}{2}
                    \biggr] \frac{\overline{f}_{2}(x)^\frac{3}{2}}{\sqrt{x}},
                    \label{eq:f2_bar_first}
            \end{eqnarray}
            
            \begin{eqnarray}
                  \frac{d^2 \overline{f}_{2}(x)}{dx^2} &=& \left[- \frac{6 \eta}{5 (5 \eta -1) \beta_{1\overline{2}}}\overline{g}_{2} 
                  -\frac{9\eta}{10(5\eta-1)\beta_{1\overline{2}}}
                   \biggl[\frac{(N_1 g_1-1)(\beta_{1\overline{2}})^\frac{3}{2}}{N_1} \right. \nonumber \\ 
                 &+& \left. \left. g_2 (\beta_{2\overline{2}})^\frac{3}{2} \right]\right]
                 \frac{\left( \overline{f}_2(x)\right)^\frac{3}{2}}{\sqrt{x}},
                 \label{eq:f2_bar_second}
            \end{eqnarray}
            
            \begin{eqnarray}
                  \frac{d^2 \overline{f}_{2}(x)}{dx^2} &=& \left[-\frac{6\eta}{5(5\eta-1) \alpha_2 \beta_{2\overline{2}}}    \overline{g}_{2} 
                  - \frac{9\eta}{10(5\eta-1) \alpha_2 \beta_{2\overline{2}}}
                  \times \biggl[\frac{(N_2 g_2-1)(\beta_{2\overline{2}})^\frac{3}{2}}{N_2} \right. \nonumber \\ 
                  &+& \left. g_1 (\beta_{1\overline{2}})^\frac{3}{2}\biggr]\right]
                  \frac{\left( \overline{f}_{2}(x) \right)^\frac{3}{2}}{\sqrt{x}}.
                \label{eq:f2_bar_third}
            \end{eqnarray}
            In a more general form, Eqs.~\eqref{eq:f2_bar_first}, \eqref{eq:f2_bar_second}, and \eqref{eq:f2_bar_third} share the same left-hand side and can therefore be combined into a single differential equation, given in Eq.~\eqref{eq:diff_eqn_f2bar_caseI}, with a constant $\overline{Q}_{2}$. The three conditions outlined in Eq.~\eqref{eq:consis_condn_f2bar_caseI}, which must be satisfied for the validity of Eq.~\eqref{eq:diff_eqn_f2bar_caseI}, will be referred to as the consistency conditions for this region.
            \begin{equation}
                   \frac{d^2\overline{f}_{2}(x)}{dx^2} = \overline{Q}_{2} \frac{\left(\overline{f}_{2}(x)\right)^\frac{3}{2}}{\sqrt{x}},
                    \label{eq:diff_eqn_f2bar_caseI}
            \end{equation}
            where,  
            \begin{eqnarray}
                \overline{Q}_{2} &=& 
                    -\frac{6\eta}{5(5\eta-1) \overline{\alpha}_2} \biggl[ g_1 (\beta_{1\overline{2}})^\frac{3}{2} + g_2 (\beta_{2\overline{2}})^\frac{3}{2}
                    \biggr] \nonumber \\
                    &=& - \frac{6 \eta}{5 (5 \eta -1) \beta_{1\overline{2}}}\overline{g}_{2} -\frac{9\eta}{10(5\eta-1)\beta_{1\overline{2}}} 
                    \left[\frac{(N_1 g_1-1)(\beta_{1\overline{2}})^\frac{3}{2}}{N_1} + g_2 (\beta_{2\overline{2}})^\frac{3}{2} \right] \nonumber \\
                   &=& -\frac{6\eta \overline{g}_{2}}{5(5\eta-1) \alpha_2 \beta_{2\overline{2}}} - \frac{9\eta}{10(5\eta-1) \alpha_2 \beta_{2\overline{2}}}
                   \biggl[\frac{(N_2 g_2-1)(\beta_{2\overline{2}})^\frac{3}{2}}{N_2}\nonumber \\ 
                   &+& g_1 (\beta_{1\overline{2}})^\frac{3}{2}\biggr].
                \label{eq:consis_condn_f2bar_caseI}
            \end{eqnarray}
            
            In the region $\overline{x}_{2} < x < x_2$, there exist two functions, $f_2(x)$ and $f_1(x)$, which are linearly related as:  
            \[
            f_1(x) = \gamma_{12} f_2(x),
            \]  
            where $\gamma_{12}$ is a linear proportionality constant. Substituting this relationship into Eq.~\eqref{eqn:tf_eqn_anticharm} yields:  
            \[
            \frac{(f_2(x))^{\frac{3}{2}}}{\sqrt{x}} = 0.
            \]  
            Since $f_2(x) \neq 0$, this equation is discarded. Subsequently, using Eqs.~\eqref{eqn:tf_eqn_charm} and \eqref{eqn:tf_eqn_light_quark}, we derive the conditions given in Eqs.~\eqref{eqn:consis_condn_midregionI} and \eqref{eqn:consis_condn_midregionII}. 
            \begin{equation}
                  \frac{d^2 f_2(x)}{dx^2}= -\frac{9\eta}{10(5\eta-1)  \gamma_{12} }\biggl[\frac{(N_1 g_1-1)(\gamma_{12})^\frac{3}{2}}{N_1} + g_2 \biggr]\frac{\left( f_2(x) \right)^\frac{3}{2}}{\sqrt{x}},
                    \label{eqn:consis_condn_midregionI}
            \end{equation}
            \begin{equation}
                 \frac{d^2 f_2(x)}{dx^2}= - \frac{9\eta}{10(5\eta-1)\alpha_2}\biggl[\frac{(N_2 g_2-1)}{N_2} 
                 + g_1 (\gamma_{12})^\frac{3}{2} \biggr]\frac{\left( f_2(x) \right)^\frac{3}{2}}{\sqrt{x}}.
                 \label{eqn:consis_condn_midregionII}
            \end{equation}
            In a more general differential equation form Eqs.~\eqref{eqn:consis_condn_midregionI},  \eqref{eqn:consis_condn_midregionII} can be written with a constant $Q_2$ as: 
            \begin{equation}
                 \frac{d^2f_2(x)}{dx^2} = Q_2 \frac{\left(f_2(x)\right)^\frac{3}{2}}{\sqrt{x}},
                  \label{eqn:consis_condn_midregion}
            \end{equation}
            where, 
            \begin{equation}
            Q_2 = \left\{
                \begin{aligned}     
                    -\frac{9\eta}{10(5\eta-1)  \gamma_{12} }\biggl[\frac{(N_1 g_1-1)(\gamma_{12})^\frac{3}{2}}{N_1} + g_2 \biggr]\\ 
                    -\frac{9\eta}{10(5\eta-1)\alpha_2}\biggl[\frac{(N_2 g_2-1)}{N_2} 
                    + g_1 (\gamma_{12})^\frac{3}{2} \biggr].
                \end{aligned}
                \right. \label{eqn:consis_condn_f2_caseI}
            \end{equation}
            The conditions given in Eq.~\eqref{eqn:consis_condn_f2_caseI}, which must be satisfied for the validity of Eq.~\eqref{eqn:consis_condn_midregion}, are referred to as the consistency conditions for this region. 

            In the region $x_2 < x < x_1$, only light quarks are present, allowing us to set $f_2(x) = 0$ and $\overline{f}_{2}(x) = 0$. Substituting this into Eqs.~\eqref{eqn:tf_eqn_anticharm} and~\eqref{eqn:tf_eqn_charm} yields:  
            \[
            \frac{(f_1(x))^{\frac{3}{2}}}{\sqrt{x}} = 0.
            \]  
            Since $f_1(x) \neq 0$, these equations are discarded. However, from Eq.~\eqref{eqn:tf_eqn_light_quark}, we obtain a differential equation of the form given in Eq.~\eqref{eqn:tf_eqn_f1_caseI}. 
            \begin{equation}
                \frac{d^2 f_1(x)}{dx^2}= -\frac{9\eta}{10(5\eta-1)}\left[\frac{(N_1 g_1-1)}{N_1} \frac{\left( f_1(x) \right)^\frac{3}{2}}{\sqrt{x}}\right].
                \label{eqn:tf_eqn_f1_caseI}
            \end{equation}
            In a more general form, Eq.~\eqref{eqn:tf_eqn_f1_caseI} can be presented as in Eq.~\eqref{eqn:diff_eqn_f1_caseI}.
            \begin{equation}
                \frac{d^2 f_1(x)}{dx^2}= Q_1 \frac{\left( f_1(x) \right)^\frac{3}{2}}{\sqrt{x}},
                \label{eqn:diff_eqn_f1_caseI}
            \end{equation}
            with, 
            \begin{equation}
                \text{}
                {Q_1} = \frac{-9\eta}{10(5\eta-1)}\frac{(N_1g_1 - 1)}{N_1}.
                 \label{eqn:consis_condn_f1_caseI}
            \end{equation}        
             In summary, the required TF equations are given by Eqs.~\eqref{eq:diff_eqn_f2bar_caseI}, \eqref{eqn:consis_condn_midregion}, and \eqref{eqn:diff_eqn_f1_caseI}. The corresponding values of $\overline{Q}_{2}$, $Q_2$, and $Q_1$ are determined from Eqs.~\eqref{eq:consis_condn_f2bar_caseI}, \eqref{eqn:consis_condn_f2_caseI}, and \eqref{eqn:consis_condn_f1_caseI}, respectively.
             
        \subsubsection{Case II: $2 \leq \eta \leq 14$}  
            In the region $0 < x < x_2$, the functions $f_1(x)$, $f_2(x)$, and $\overline{f}_{2}(x)$ are present and are related by the expressions:  
            \[
            f_1(x) = \beta_{12} f_2(x), \quad \overline{f}_{2}(x) = \beta_{\overline{2}2} f_2(x),
            \]  
            where $\beta_{12}$ and $\beta_{\overline{2}2}$ are linear proportionality constants. By following a similar procedure as in Case I, we obtain an equation of the form of Eq.~\eqref{eqn:consis_condn_midregion} with new $Q_2$ given by: 
            
            \begin{equation}
                Q_{2}  =  \left\{ \begin{aligned}
                            &  - \frac{6 \eta}{5 (5\eta -1)\overline{\alpha}_{2} \beta_{\overline{2}2}} \biggl[ g_{1} (\beta_{12})^\frac{3}{2} + g_{2} \biggr] \\ & 
                             - \frac{6 \eta}{5 (5 \eta -1) \beta_{12}} \overline{g}_{2} (\beta_{\overline{2}2})^\frac{3}{2} - \frac{9 \eta}{10 (5\eta-1) \beta_{12}} \left[ \frac{(N_{1}g_{1} -1)}{N_{1}}(\beta_{12})^\frac{3}{2} + g_{2} \right] \\ &
                             - \frac{6 \eta}{5 (5 \eta -1) \alpha_{2}} \overline{g}_{2} (\beta_{\overline{2}2})^\frac{3}{2} - \frac{9 \eta}{10 (5 \eta -1 ) \alpha_{2}} \biggl[ \frac{(N_{2}g_{2} -1)}{N_{2}} + g_{1} (\beta_{12})^\frac{3}{2} \biggr].
                        \end{aligned}
                \right.
                \label{eqn:consis_condn_f2_caseII}
            \end{equation}
            Similarly, in the region $x_2 < x < \overline{x}_{2}$, the functions $\overline{f}_{2}(x)$ and $f_1(x)$ are present and are related by:  
            \[
            f_1(x) = \gamma_{1\overline{2}} \overline{f}_{2}(x),
            \]  
            where $\gamma_{1\overline{2}}$ is a proportionality constant. Using this relationship, we derive an equation of the form of Eq.~\eqref{eq:diff_eqn_f2bar_caseI} with new $\overline{Q}_2$ given by: 
            \begin{equation}
                \overline{Q}_{2} = \left\{ 
                \begin{aligned}
                    &  - \frac{6 \eta}{5 (5 \eta -1) \overline{\alpha}_{2}} g_{1} (\gamma_{1\overline{2}})^\frac{3}{2}\\ 
                    &   -\frac{6\eta \overline{g}_{2}}{5(5\eta-1) \gamma_{1\overline{2}}} - \frac{9 \eta (N_{1}g_{1} -1)}{10 (5\eta -1)N_{1}} (\gamma_{1\overline{2}})^\frac{1}{2}.
                \end{aligned}
                \right. \label{eqn:consis_condn_f2bar_caseII}
            \end{equation}
            In the region $\overline{x}_{2} < x \leq x_1$, only the function $f_1(x)$ exists, while the remaining functions satisfy $f_2(x) = 0$ and $\overline{f}_{2}(x) = 0$. This leads to the same differential equation, Eq.~\eqref{eqn:diff_eqn_f1_caseI}, along with the corresponding consistency condition given by Eq.~\eqref{eqn:consis_condn_f1_caseI}, as in Case I.
           \begin{table*}[h!]
            \small
            \caption{Parameters obtained by solving consistency conditions. The proportionality constants for Case I are $\beta_{1\overline{2}}$, $\beta_{2\overline{2}}$, and $\gamma_{12}$, whereas for Case II they are $\beta_{12}$, $\beta_{\overline{2}2}$, and $\gamma_{1\overline{2}}$.
            }
               \begin{center}
            \begin{tabular}{cccccccccc}
                \hline
                $\eta$ & $\beta_{12}/\beta_{1\overline{2}}$ & $\beta_{2\overline{2}}/\beta_{\overline{2}2}$ & $\gamma_{12}/\gamma_{1\overline{2}}$  & $N_{1}$ & $g_{1}$ & $N_2$ & $g_2$ & $\overline{N}_{2}$ & $\overline{g}_{2}$\\
                \hline 
                1 & 0.27530 &  0.96216 & 0.39542& 1 & 3& 1& 1&1&1\\
                2 & 0.24858 &0.88262 & 0.49745 & 2        & 3   & 1&2&1&2 \\
                3 & 0.22560  &0.85936 & 0.39542 & 3 & 3& 3& 1&3&1 \\
                4 & 0.22047 & 0.82371& 0.45567 & 3 & 4&2&2&2&2 \\
                5 & 0.21394 & 0.82499& 0.39872& 5 & 3& 5& 1&5&1 \\
                6 & 0.21321 &0.79469  & 0.50524 & 6 & 3&3&2&3&2\\
                7 & 0.20902 & 0.81049& 0.40014 & 7 & 3& 7& 1&7&1 \\
                8 & 0.20906 & 0.78413   & 0.50623 & 6 & 4&4&2&4&2 \\
                9 & 0.20632 & 0.80251   & 0.40094 & 9 & 3& 9& 1&9&1\\
                10 & 0.20660 & 0.77785   & 0.50682 & 10 & 3&5&2&5&2 \\
                11 & 0.20460 & 0.79745   & 0.40144 & 11 & 3&11&1&11&1 \\
                12 & 0.20462 & 0.78127   & 0.45969 & 9 & 4&6&2&6&2 \\
                13 & 0.20343 & 0.79396   & 0.40179 & 13 & 3&13&1&13&1 \\
                14 & 0.20382 & 0.77072   & 0.50750 & 14 & 3&7&2&7&2 \\
                \hline
            \end{tabular}
            \end{center}
            \label{tab:parameter_values}
        \end{table*}  
        We solved the consistency conditions described above, using \textbf{NDSolve} from {\it Mathematica}, for both cases to obtain the parameters for up to 70 quarks. The procedure for solving these conditions is described in further detail in Section~\ref{sec:computational_details}. 
        
        Interestingly, these conditions in the innermost region resulted in five different solutions, one with the real values of the parameters and the others with the complex values. These multiple solutions occur because of the presence of a fifth-order power of the parameter in the equation. Similarly, the middle region, with third-order power of the parameter in the consistency condition, resulted in three solutions, one with real parameters and two with complex parameters. In addition, in the outermost region, the consistency condition resulted in a single solution with the real parameters. It is remarkable that the model gives only one set of real solutions, which increases our confidence in its consistency.  
   
        The results of the solution with the real parameters are presented in Table~\eqref{tab:parameter_values}. 
        Note that the proportionality constants for Case I ($\eta = 1$) are $\beta_{12}$, $\beta_{2\overline{2}}$ and   $\gamma_{12}$, whereas for Case II ($\eta = 2-14$) they are $\beta_{1\overline{2}}$, $\beta_{\overline{2}2}$, and $\gamma_{1\overline{2}}$. With the introduction of two separate regions for the heavy quarks and one region for the light quarks the parameters presented in Table~\eqref{tab:parameter_values} differ from those in Table 10 from~\cite{giri2021investigation}. This difference arises from a change in notation in the present work. The proportionality constants in the relationship between the TF functions for Case I are now expressed as $f_1(x) = \beta_{1\overline{2}} \overline{f}_{2}(x)$ and $f_2(x) = \beta_{2\overline{2}} \overline{f}_{2}(x)$, compared to the earlier notation $f_1(x) = k f_2(x)$ and $\overline{f}_{2}(x) = \overline{k} f_2(x)$. This results in the relation $k = \beta_{1\overline{2}} \text{/} \beta_{2\overline{2}}$ and $\overline{k} = 1 \text{/} \beta_{2\overline{2}}$. 
        For the Case II, the parameter $k$ is replaced by $\beta_{12}$ while $\overline{k}$ is replaced by $\beta_{\overline{2}2}$. Parameters from both of these tables agree up to the fourth decimal place. 

    \subsection{Total Energy}    
    \label{subsection:total_energy}

    We are interested in investigating family stability questions within this model. Thus, we investigate the most degenerate TF ground states, which will be the most stable for a given $\eta$ pocket. These values are specified in Table~\ref{tab:parameter_values}. For example, for $\eta=1$ we put all three light quarks in the ground state with $g_1=3$. For $\eta=2$, we can max out the heavy quark degeneracies with $g_2=2$ and $\bar{g}_2=2$. A special case is $\eta=4$ and integer multiples of this value, where we max out the degeneracies of both the light and heavy quark sectors with $g_1=4$ and $g_2=\bar{g}_2=2$. As we will see, these are an especially stable type of state. We will have much more to say about these configurations later. Table~\ref{tab:parameter_values} is thus completed up to $\eta=14$.
    
        The total energy of the system is obtained by adding the kinetic energies arising from the motion of the particles, the potential energies resulting from the interactions between the particles, and the volume energies associated with the confinement of the system. The expression for the kinetic energy of the system can be obtained by solving Eq.~\eqref{eqn:kinetic_energy}. Here, $I$ denotes the flavor of quarks and $\alpha_{s}$ the strong coupling constant. $r_{max}$ is the integration limit that defines the boundary of different regions and differentiates the two cases of our interest. 
         \begin{equation}
            \begin{split}
                K.E.  &= \sum_I \int^{r_{max}} d^3r \frac{3(6\pi^2\hbar^3)^\frac{5}{3}}{20\pi^2\hbar^3m^I(g^I)^\frac{2}{3}} \biggl[\biggl[\frac{2 \times (4/3)\alpha_s}{a}\biggr]^\frac{3}{2} \biggl(\frac{f^I(r)}{r}\biggr)^\frac{3}{2} \frac{g^I}{6\pi^2} \biggr]^\frac{5}{3} \\ & \quad + \sum_I \int^{r_{max}} d^3r \frac{3(6\pi^2\hbar^3)^\frac{5}{3}}{20\pi^2\hbar^3\overline{m}^I (\overline{g^I})^\frac{2}{3}} \biggl[\biggl[\frac{2 \times (4/3)\alpha_s}{a}\biggr]^\frac{3}{2} \biggl(\frac{\overline{f}^I(r)}{r}\biggr)^\frac{3}{2} \frac{\overline{g}^I}{6\pi^2}\biggr]^\frac{5}{3}.
                \label{eqn:kinetic_energy}
            \end{split}
        \end{equation}        
        Eq.~\eqref{eqn:expression_of_potential_energy} gives the potential energy of the system for both cases. The parameters $K_{11}$, $K_{22}$, $K_{\overline{2}\overline{2}}$, $K_{12}$, $K_{21}$, $K_{\overline{2}1}$, and $K_{\overline{2}2}$, are obtained by solving the sum of two double integrals. The analytical expressions for these parameters are given in~\ref{appendix I} and~\ref{appendix II}. Details of the calculations with individual terms required to get the total energy for both cases can be found in~\cite{aryal2024energy}.
        \begin{eqnarray}
            P.E. &=& -C\left[\frac{(g_1 N_1 - 1)}{N_1} g_1 K_{11}
            + \frac{(g_2 N_2 - 1)}{N_2} g_2 K_{22}
            + \frac{(\overline{g}_{2} \overline{N}_{2} - 1)}{\overline{N}_{2}} \overline{g}_{2} K_{\overline{2}\overline{2}} \right. \nonumber \\ 
            &+& g_{1} g_{2} K_{12} + g_{2} g_{1} K_{21}
            + \left. \tfrac{4}{3} \overline{g}_{2} g_1 K_{\overline{2}1} + \tfrac{4}{3} \overline{g}_{2} g_2 K_{\overline{2}2}\right].
            \label{eqn:expression_of_potential_energy}
        \end{eqnarray}
        Furthermore, the volume energy of the system is $\frac{4}{3} \pi R^{3} x_{1}^3 B$, with $B = 107.6$ MeV~\cite{Baral:2019qpl}. With the total energy and the parameters obtained by solving the consistency conditions, we explore the properties by analyzing the energy trends of these systems. The computational algorithm used to solve the TF equations is further
        elucidated in the Section 3.

\section{Computational Details}
    \label{sec:computational_details}
        The {\it Mathematica} program is used to solve the consistency conditions for different $\eta$ values. At first, the value of constants like coupling constant, degeneracy level, and the number of quarks in that degeneracy level is defined. Then,  using the built-in \textbf{NSolve} function we solve Eqs.~\eqref{eq:consis_condn_f2bar_caseI},~\eqref{eqn:consis_condn_f2_caseI}, ~\eqref{eqn:consis_condn_f2_caseII}, and ~\eqref{eqn:consis_condn_f2bar_caseII} to find the numerical values of parameters from Table~\eqref{tab:parameter_values}. Global variables that were used for the optimization in {\it Mathematica} are shown in Table~\eqref{tab:optimization_parameters}~\cite{Baral:2019qpl}.
       The \textbf{NDSolve} function numerically solves the differential equations within the given boundary; the code is then initialized by a guessed choice of the first boundary value and iterates until a minimum energy value is found. By updating the approximate solutions with the use of nested loops, the objective function is optimized until it reaches a local minimum. For that purpose, the \textbf{FindMinimum} function is utilized, which employs a gradient descent algorithm and iteratively adjusts the parameters, based on gradient information, until a convergence criterion is met. 
       \begin{table}[h!]
    \caption{The parameters leveraged during the optimization process of the code. The value of these constants is taken from~\cite{Baral:2019qpl}. The other constants that are found in the code are merely the ratio of parameters shown here.}
       \begin{center}
             \begin{tabular}{c c  }
                        \hline
                        Global Parameters & Values  \\
                        \hline
                        Reduced Planck's constant & 1 \\
                        Velocity of light & 1 \\
                        Strong coupling constant & 0.346 \\
                        Mass of light quark & 306 MeV \\
                        Mass of charm quark & 1553 MeV \\
                        Bag constant & 107.6 MeV \\
                        \hline
                        \end{tabular}
                           \end{center}
                \label{tab:optimization_parameters}        
                \end{table}
        
\section{Results and Discussions}
\begin{figure*}
        \centering
        \includegraphics[width = \textwidth]{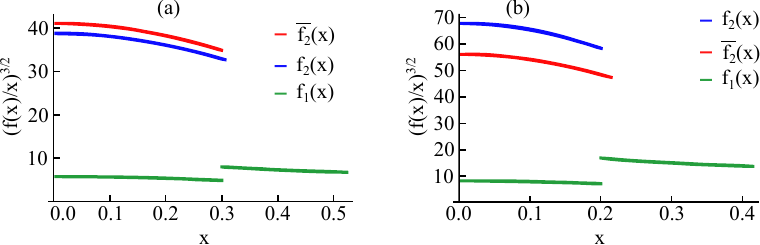}
        \caption{The density function plot for (a) $\eta=1$ and (b) $\eta=2$ as a function of distance from the center.}
        \label{fig:density_function_plot}
    \end{figure*}

    Figure \ref{fig:density_function_plot}(a) shows the density functions of quarks with respect to the dimensionless parameter \textit{x} for a single pocket of pentaquark, considered as Case I. It is seen that the density functions of pentaquarks decrease with the increase in \textit{x} initially. The anti-charm quark, depicted with the red line, is confined inside the inner region only. The blue curve denotes the charm quark and it is continuous across the boundary. Furthermore, the light quark, denoted by the green line, shows a discontinuity at the boundary of the inner region. The discontinuity shows that the density function of the light quark increases abruptly going on to the middle region. Note that light quark has a very low density in the inner region but increases abruptly. Then it continues to decrease until it reaches the outermost boundary of the pentaquark system.      
         
    The nature of the plot of density functions for Case II is shown in Fig \ref{fig:density_function_plot}(b).  The figure is actually for double pockets of pentaquarks but the nature of the graph is similar for all higher pockets of pentaquarks. The difference in plots of density functions between the two cases is that $\overline{f}_{2}(x)$ lies in the inner region for Case I whereas for Case II, $f_{2}(x)$ lies in that region.    

    \begin{table}[h!]
    \caption{Comparison of energy per quark when $f_{1}(x)$ is continuous and discontinuous. $f_{1}(x)$ is made continuous by assuming $\beta_{1\overline{2}} = \beta_{2\overline{2}} \gamma_{12}$ for Case I and  $\beta_{12} = \beta_{\overline{2}2} \gamma_{1\overline{2}}$ for Case II at the boundary between inner and middle region. The energies are in MeV.}
     \begin{center}
        \begin{tabular}{ c c c c }
            \hline
            \small{$\eta$} & \small{No. of quarks} & \small{Discontinuous} & \small{Continuous} \\
            \hline
            1 & 5  & $34.55$ &   $36.31$     \\
            2 & 10  & $35.12$ &  $38.20$    \\
            3 & 15  & $37.95$ & $40.25$    \\
            4 & 20  & $32.56$ & $35.21$    \\
            5 & 25  &  $39.18$  & $41.62$     \\
            6 & 30  & $37.91$ & $41.44$\\
            7 & 35  & $39.87$ & $42.39$    \\
            8 & 40  & $34.09$ & $36.90$     \\
            9 & 45  & $40.34$ & $42.90$     \\
            10 & 50  & $38.90$  & $42.54$     \\
            11 & 55  & $40.69$  & $43.27$     \\
            12 & 60  & $34.83$  & $37.70$     \\
            13 & 65  & $40.96$  & $43.56$     \\
            14 & 70  & $39.45$  & $43.15$     \\
            \hline
                \label{tab:energy_comparison}     
        \end{tabular}
          \end{center}
        \end{table}
     As explained earlier, our system has three distinct regions where three functions exist for the innermost region, two for the middle region and one for the outermost region. TF model assumes the continuity of the function and its derivatives at the boundary. This led to an interesting scenario. Out of the two TF functions at the intermediate region, we could only make one of them continuous at the boundary of innermost and middle region. Making one continuous automatically makes another discontinuous. So, we calculated the system energies for both the cases as shown in Table~\eqref{tab:energy_comparison}. In this table, we can see that if we make $f_1(x)$ continuous, the energy actually ends up being greater than it being discontinuous. Since TF function is used to calculate the ground state energy, all the analysis done hereafter shows a discontinuity of the light quark in the inner boundary. Note that $f_{1}(x)$ is made continuous by assuming $\beta_{1\overline{2}} = \beta_{2\overline{2}} \gamma_{12}$ for Case I and  $\beta_{12} = \beta_{\overline{2}2} \gamma_{1\overline{2}}$ for Case II at the boundary between inner and middle region as our code automatically assumes $f_{2}(x)$ to be continuous for $\eta = 1$ and $\bar{f}_{2}(x)$ to be continuous for $\eta > 1$.
    
    \begin{figure*}
        \centering
        \includegraphics[]{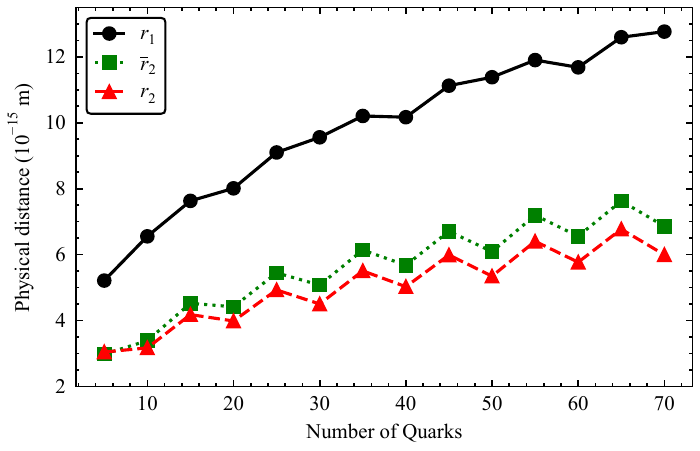}
        \caption{The black line shows physical distance (in $fm$) versus quark content, indicating the increase in distance with the increase in quark content. The green and red lines refer to the boundaries of anti-charm and charm, respectively. }
        \label{fig:physical_distance_discontinuous}
    \end{figure*}     
    In Figure~\eqref{fig:physical_distance_discontinuous}, the physical radius is plotted versus the quark number. The physical radius for the anti-charm quark is greater than the radius for the charm quark except at $\eta=1$ which is as expected. The physical distance is the product of constant $R$ and the dimensionless radius ${x}$. Tendency of all the dimensional radii is to decrease, as seen in Table \ref{tab:physical_radius_comparison}, whereas ${\eta^\frac{2}{3}}$ in ${R}$ increases. Therefore, the result is an increase in overall radius with increasing quark content. Also, the curve of the outermost radius plot tends to flatten out for a large number of quarks. The zig-zag (zig-zag in the sense of increasing and decreasing continuously) nature of the plot for the inner boundaries is due to the presence of the degeneracy factor. Degeneracy allows quarks to stack up which decreases the distances of the inner boundaries.       

    \begin{table}[h!]
        \caption{Comparison of dimensionless parameters that determine the physical radius of the system. The value of $\overline{x}_{2}$ is less than $x_{2}$ only for $\eta = 1$.}
        \begin{center}
            \begin{tabular}{c c c c c  c }
            \hline
            $\eta$   & $\overline{x}_{2}$  & & $x_{2}$ & & $x_{1}$ \\
            \hline
            1    & 0.29923 & &  0.30547  & & 0.52337  \\
            2    & 0.21422 & &  0.20042  & & 0.41469  \\
            3    & 0.21835 & & 0.20171   & & 0.36845 \\
            4    & 0.17594 & & 0.15882   & & 0.31930 \\
            5    &  0.18705 & & 0.16926   & &  0.31258 \\
            6    & 0.15437 & & 0.13688   & &  0.29080\\
            7    & 0.16847 & & 0.15112   & & 0.28019 \\
            8    & 0.14135 & & 0.12452   & &  0.26503 \\
            9    & 0.15570 & & 0.13899   & &  0.25818 \\
            10    & 0.13180 & & 0.11567   & &  0.24635 \\
            11    & 0.14610 & & 0.13003   & &  0.24175 \\
            12    & 0.12565 & & 0.11054   & &  0.22390 \\
            13    & 0.13853 & & 0.12304   & &  0.22887 \\
            14    & 0.11861 & & 0.10366   & &  0.22079 \\
            \hline
            \end{tabular}
            \end{center}
            \label{tab:physical_radius_comparison}        
    \end{table}  
    
    \begin{figure*}
        \centering
        \includegraphics[]{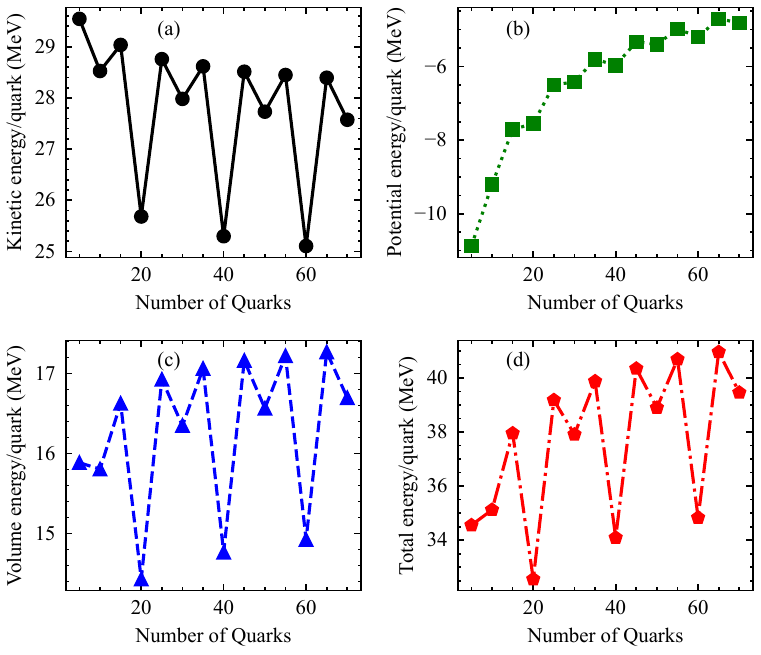}
        \caption{The plot of (a) kinetic energy (b) potential energy (c) volume energy, (d) total energy per quark (in MeV) without rest mass energy for the case of $\eta = 1\text{-}14$. }
        \label{fig:energy_plot_discontinuous}
    \end{figure*}

    \begingroup
       \renewcommand{\thefootnote}{\fnsymbol{footnote}}
        Within this model, we can identify three distinct forms of energy: kinetic energy, potential energy, and volume energy. As seen in Figure \ref{fig:energy_plot_discontinuous}(a) the kinetic energy per quark keeps on decreasing and increasing in a zig-zag pattern and there is a sudden drop for $\eta=4$, and again for $\eta=8$ and 12. Figure \ref{fig:energy_plot_discontinuous}(b) shows that the potential energy per quark increases up to $\eta=3$ increases sharply but beyond that, the energy increase seesaws with energy first increasing in small amounts and then in large amounts. The nature of the graph for volume energy per quark can be seen in Figure \ref{fig:energy_plot_discontinuous}(c). We can observe dips in energies for even number of quarks and rise in energies for odd number of quarks. That happens because the degeneracy is high for even number of quarks whereas it is low for odd number of quarks. Most importantly, there is again a sharp decrease in energy for $\eta=4, 8, 12$.  Higher degeneracy allows the radius to decrease and hence the decrease in volume energy.
        Figure~\ref{fig:energy_plot_discontinuous}(d) is the plot of total energy per quark i.e., the sum of kinetic, potential, and volume energy, against the quark content.
        It indicates that the total energy per quark reaches a minimum for four pockets of pentaquarks, although there are similar decreases for $\eta= 8$ and 12. The lowest overall energy per quark is for $\eta=4$, implying that this configuration is the most stable among pentaquark TF families\footnote[1]{We have reserved designation for this 20-quark state as the Icosaquark Thunderwolf (TW-20), in tribute to SUNY Niagara County Community College’s Thunderwolves.}. 
    \endgroup
            
\section{Interpretation of Results}
    Figure~\ref{fig:energy_plot_discontinuous}(d) for $\eta=4$ (20 quarks) is a remarkable result, but must be interpreted carefully. What is shown is the energy state of the most degenerate combination of light, $g_1=4, N_1=3$, and heavy quarks, $N_2=2,g_2=2,\bar{N}_2=2,\bar{g}_2=2$ (see Table~\ref{tab:parameter_values}). This would be considered a state if the individual spins were considered good quantum numbers. Of course, this is not what is expected.
    
    Let us recall the procedure for assigning particle states in the TF quark model. The TF quark model is a semi-classical method which is based on weighted probabilities of configurations rather than actual quantum amplitudes. However, the probabilities are based upon the quantum wave function. In Ref.~\cite{Liu:2012my}, we considered the example of the appropriately symmetrized non-relativistic proton flavor-spin $j=\frac{1}{2}$, $m=\frac{1}{2}$ wave function:
    \begin{eqnarray}
        |P,+\rangle &\equiv& |uud\rangle \left(2|++-\rangle-|+-+\rangle
        - |-++\rangle\right)/(3\sqrt{2}) \nonumber \\
        &+& {\rm cyclic\, permutations}.
        \label{proton}
    \end{eqnarray}
    Using the flavor degeneracy factors, these two configurations are classified as:
    \begin{eqnarray*}
    & (u^{\uparrow}u^{\uparrow}d^{\downarrow}):& \, g_1=1, N_1=2;\, g_2=1, N_2=1.\\ 
    & (u^{\uparrow}u^{\downarrow}d^{\uparrow}):& \, g_1=3, N_1=1; N_2=0.  \label{pconfig3}
    \end{eqnarray*}
    (There is no significance to the flavor or spin sequential ordering of these $uud$ configurations.) That is, the first configuration has two identical particles (the $u$ quarks) and a second generalized flavor ($d$), whereas the second configuration simply has three non-identical but mass-degenerate quarks, with no second set of particle labelings necessary. The proton is then said to be in the TF configuration:
    \begin{eqnarray*}
    \frac{2}{3}(u^{\uparrow}u^{\uparrow}d^{\downarrow})+\frac{1}{3}(u^{\uparrow}u^{\downarrow}d^{\uparrow}).   \label{pconfig2}
    \end{eqnarray*}
    The mass of this state is the weighted average of these two calculable TF configurations.
    
    We propose to treat the spin state of the multi-pentaquark system in a similar manner. However, the states we are considering are similar to atomic states containing a heavy nucleus and light electrons. The spin states of these systems are considered separately and good spin quantum numbers are assigned to each. Although our situation is not as extreme, we propose to consider the total spin of the light and heavy quarks also as separate good quantum numbers, which can then be combined to a final spin angular momentum quantum state. 
    
    Let us take the simpler case of $\eta=2$ to discuss the construction and interpretation of the physical states. This would involve 6 light $u$ or $d$ quarks as well as two charm and two anti-charm quarks. Just like the lowest energy charmonium state has spin 0, it is clear that the lowest energy heavy quark spin state is a double-charmonium spin state 0. Table~\ref{tab:parameter_values} shows the light quark TF configuration being evaluated for $\eta=2$ has $g_1=3$ and $N_1=2$. This configuration has spin magnetic quantum number of $m_z=\pm 1$, and the lowest mass manifestation most probably corresponds to a spin 1 state. Restricting ourselves to the positive charge 2 and $m_z=1$ version, it may be characterized by the TF formation:
    \begin{equation*}
    (u^{\uparrow}u^{\downarrow}d^{\uparrow})(u^{\uparrow}u^{\downarrow}d^{\uparrow}): \, g_1=3, N_1=2.
    \end{equation*}
    Here we are applying a maximal degeneracy rule for assigning the configuration. Designating this state as \lq\lq $\eta=2$", the wave function of this has the general form:
    \begin{eqnarray}
        &|\eta=2,+\rangle\equiv |uuuudd\rangle \left(C_1|+-+-++\rangle+\text{ 15 additional terms}\right) \nonumber\\&+{\rm cyclic\, permutations}.\label{proton}
    \end{eqnarray}
    It is clear from this form that other TF configurations will be part of this wavefunction, including:
    \begin{eqnarray*}
        & (u^{\uparrow}u^{\downarrow}d^{\uparrow}d^{\downarrow})(u^{\uparrow}u^{\uparrow}): \, g_1=4, N_1=1;\, g_{1'}=1, N_{1'}=2,\\ 
        & (u^{\uparrow}d^{\downarrow})(u^{\uparrow}d^{\downarrow})(u^{\uparrow}u^{\uparrow}): \, g_1=2, N_1=2; g_{1'}=1,N_{1'}=2. 
    \end{eqnarray*}
    The introduction of the additional $1'$ particle label makes it clear that in general we will encounter mixed degeneracy states, like the proton. There it was possible to include such configurations because two physical regions were considered in the calculation. In our present case, this would require the construction of a model with four physical regions: two for the charm and anti-charm quarks, and two for the different configurations of light quarks.
    
    There are thus two barriers to our construction of the lowest mass physical state associated with $\eta=2$. There is a combinatorial problem associated with the construction of the appropriate wave functions, as well as the limitation of our present 3-region analysis model to accommodate and calculate the necessary configurations. The $\eta=4$ case we have evaluated has $g_1=4$ and $N_1=3$. This configuration has spin magnetic quantum number $m_z=0$. It also has charge 2 and the lowest mass manifestation most probably corresponds to a spin 0 state. The combinatorial and calculational issues seen for $\eta=2$ will continue for the higher $\eta$ states. These problems are not insurmountable, but will require further research and analysis before the exact states involved can be identified.
    

\section{Summary and Conclusions}

The Thomas-Fermi quark model is an effective theoretical tool which can reveal systematics of many-quark states. This is very difficult for lattice QCD calculations, which isolate one particular state at a time. In this paper we continue our investigation of hidden charm multi-pentaquark families using the TF quark model, which requires three separate spherical regions. Our initial setup of the problem in~\cite{giri2021investigation} determined the interaction weights, various color interaction terms as well as the probabilities of particle and antiparticle interactions. This allowed us to form the kinetic and potential energies as well as the appropriate TF differential equations and consistency conditions. Using an iterative method we have explicitly evaluated the functional forms, radial boundaries and associated energies for pockets of up to 14 pentaquarks. The bound state of a single pentaquark was found to have a different quark radial structure compared to multi-pentaquarks; the single pentaquark state had the anti-charm content fully contained in the innermost region, whereas multi-pentaquarks had the charm quark content fully contained in this region. When the energies for states were determined, we also discovered that configurations with a discontinuity in the light quark $f_1$ function at the inner radius ($\bar{x}_2$ for the $\eta = 1$ case and $x_2$ for $\eta> 1$) were energetically preferred over continuous $f_1$ solutions.

A picture emerges of the growth of the systems as the number of pockets is increased, as seen in Figure~\ref{fig:physical_distance_discontinuous}, which is plotted for the maximally degenerate ground states. The outer radius increases almost smoothly whereas the two inner radii show a zig-zag pattern due to the presence of the degeneracy factor for the charm quarks. Note that this outer radius is significantly larger than was found for an equivalent number of particles in our multi-quark meson investigation involving charm and bottom quarks~\cite{Baral:2019qpl}. 

There are three sources of energy in our model: kinetic, potential and volume energy. The numerical results for pockets of up to 14 pentaquarks were displayed in the various panels of Figure~\ref{fig:energy_plot_discontinuous}. The rest mass is not included in these results and the energies are divided by the number of quarks in order to evaluate the stability. States with a lower total energy have a greater quark binding energy and are more stable. Our results show that there is a dramatic decrease in the total energy when one considers the pentaquark family with integral multiples of $\eta=4$ pockets. These do not appear to be a stable family as the total energy still increases as $\eta$ increases. Instead, we find that the total energy per quark is minimum for a $\eta=4$ pocket of pentaquarks, which we term an icosaquark, implying it is the most stable TF pentaquark state. As pointed out in Section~5, the angular momentum content and other properties associated with this state must still be determined through the special rules developed in Ref.~\cite{Liu:2012my}. There are a tower of states which can be built up from 12 light quarks and 8 heavy quarks. We will take up this question in our further work.

Note that there are many other types of quark systems which can be explored with this approach. One of these would be pentaquarks including a strange quark, another would be multi-baryon states with heavy quarks, as well as mixed baryon-mesonic states. A refinement that needs to be incorporated in these studies is the inclusion of spin energy~\cite{Liu:2012my}.

\section{Acknowledgments}
We thank the Baylor University Research Committee and the Baylor Graduate School for their partial support.
We would like to acknowledge Mr. Sujan Baral and Ms. Pratigya Gyawali, CEO and COO, respectively, of the Everest Institute of Science and Technology, for initiating the EVIST research collaborations.
We also acknowledge the Grant Office at SUNY Niagara, New York, for their assistance.
Special thanks go to Nirmal Dangi and Dipu Nepal for their contributions to partial calculations in this paper.
We also extend our gratitude to several scholars from SUNY Niagara who participated in related Thomas–Fermi projects at various times, namely Elijah Alker, Zachary Farrell, Evan O’Connor, and Jacob Mongold.

\appendix
\section{} \label{appendix I}
The expression of kinetic energy for Case I is 
\begin{equation}
 \begin{split}
 K.E_{\text{case }I} &=
 C \biggl[
 \frac{4}{7} 
    \biggl[\alpha_1g_1
        \biggl[(\beta_{1\overline{2}})^\frac{5}{2} 
        - (\gamma_{12})^\frac{5}{2} 
            \left(\frac{\beta_{1\overline{2}}}{\gamma_{12}}
            \right)^\frac{5}{2} 
        \biggr]
       + \overline{\alpha_2}\overline{g}_{2}
        \biggr] \\ & \times \sqrt{x} \left. (\overline{f}_{2}(x))^\frac{5}{2} \right|_{\overline{x}_{2}}  + \left. \frac{4}{7} \alpha_2g_2  \sqrt{x} (f_2(x))^\frac{5}{2} \right|_{x_2} 
        +\left. \frac{4}{7} \alpha_1g_1 \sqrt{x} (f_1(x))^\frac{5}{2}\right|_{x_1}  \\ &
        + \frac{5\overline{N}_{2}}{21\eta}         
        \biggl[
            \left(\frac{\beta_{1\overline{2}}}{\gamma_{12}}\right)^2   \frac{\overline{Q}_{2}}{Q_2}\biggl(\alpha_1g_1(\gamma_{12})^\frac{5}{2} 
            + \alpha_2g_2\biggr) - \alpha_1g_1 (\beta_{1\overline{2}})^\frac{5}{2}  \\ & -\alpha_2g_2 (\beta_{2\overline{2}})^\frac{5}{2}   - \overline{\alpha_2}\overline{g}_{2} \left. 
        \biggr]
            \frac{d\overline{f}_{2}(x)}{dx}\right|_{\overline{x}_{2}}
            + \frac{5}{21\eta} \biggl[
            \frac{(\gamma_{12})^2 \alpha_{1}g_{1}}{Q_1} \biggl[N_2Q_2 \\ & +\left(\frac{\beta_{1\overline{2}}}{\gamma_{12}}\right)\overline{N}_{2}\overline{Q}_{2}  - (\beta_{2\overline{2}})^\frac{3}{2} \overline{N}_{2}Q_2    
        \biggr]  -  
        \biggl[        
             \alpha_1g_1(\gamma_{12})^\frac{5}{2} + \alpha_2g_2\biggr]  \biggl[N_2 \\ & +\left(\frac{\beta_{1\overline{2}}}{\gamma_{12}}\right)\frac{\overline{N}_{2}\overline{Q}_{2}}{Q_2} - (\beta_{2\overline{2}})^\frac{3}{2}\overline{N}_{2}\biggr]\left.
        \biggr]  \frac{df_2(x)}{dx}\right|_{x_2} - \frac{5\alpha_1g_1}{21\eta}             
        \biggl[ N_1 \\ & + \gamma_{12}\frac{N_2Q_2}{Q_1}    + (\gamma_{12})^\frac{3}{2}  \biggl((\beta_{2\overline{2}})^\frac{3}{2}\overline{N}_{2}  - N_2\biggr) \\ & - (\beta_{1\overline{2}})^\frac{3}{2}\overline{N}_{2}   - (\gamma_{12})(\beta_{2\overline{2}})^\frac{3}{2} \frac{\overline{N}_{2}Q_2}{Q_1} \\ & + (\gamma_{12})\left(\frac{\beta_{1\overline{2}}}{\gamma_{12}}\right)  \frac{\overline{N}_{2}\overline{Q}_{2}}{Q_1}\left. \biggr]\frac{df_1(x)}{dx} \right|_{x_1}\biggr],           
  \label{eqn:kinetic_energy_caseI}
  \end{split}
  \end{equation}
            The final expressions of constants that appear in the potential energy term for Case I are shown below.  
            \begin{equation}
               \begin{split}
              K_{11} &= \frac{4}{7} \biggl[ \frac{ \left(\frac{\beta_{1\overline{2}}}{\gamma_{12}}\right)^\frac{5}{2} (\gamma_{12})^3}{Q_2} -  \frac{(\beta_{1\overline{2}})^3}{\overline{Q}_{2}} \biggr] \sqrt{x} \left. (\overline{f}_{2}(x))^\frac{5}{2} \right|_{\overline{x}_{2}} \\&
              +  \frac{4}{7} \biggl[\frac{(\gamma_{12})^\frac{5}{2}}{Q_1} - \frac{(\gamma_{12})^3}{Q_2} \biggr]  \left. \sqrt{x} (f_2(x))^\frac{5}{2} \right|_{x_2}   -\left. \frac{4}{7 Q_1} \sqrt{x} (f_1(x))^\frac{5}{2} \right|_{x_1}  \\ & +
          \Biggl[
              \frac{\overline{N}_{2}}{3 \eta} \Biggl( \frac{(\beta_{1\overline{2}})^3}{\overline{Q}_{2}}  - \frac{(\beta_{1\overline{2}})^\frac{3}{2} (\gamma_{12})^\frac{3}{2}  \left(\frac{\beta_{1\overline{2}}}{\gamma_{12}}\right)}{Q_2}  + \frac{\overline{Q}_{2}}{{Q_2}^2} (\gamma_{12})^3  \left(\frac{\beta_{1\overline{2}}}{\gamma_{12}}\right)^2  \\ & - \frac{(\beta_{1\overline{2}})^\frac{3}{2} (\gamma_{12})^\frac{3}{2}  \left(\frac{\beta_{1\overline{2}}}{\gamma_{12}}\right)}{Q_2} \Biggr) + \frac{5 \overline{N}_{2}}{21 \eta} 
              \Biggl(
              \frac{(\beta_{1\overline{2}})^3}{\overline{Q}_{2}}   
              - \left. \frac{\overline{Q}_{2}}{{Q_2}^2}  \left(\frac{\beta_{1\overline{2}}}{\gamma_{12}}\right)^2 (\gamma_{12})^3 \Biggr) \Biggr] \frac{d\overline{f}_{2}(x)}{dx} \right|_{\overline{x}_{2}}  \\ & + 
              \Biggl[ \frac{\overline{N}_{2}}{3 \eta}  \Biggl( \frac{(\beta_{1\overline{2}})^\frac{3}{2} (\gamma_{12})^\frac{3}{2}}{Q_2}  - \frac{(\beta_{1\overline{2}})^\frac{3}{2} \gamma_{12}}{Q_1}  - \frac{(\beta_{2\overline{2}})^\frac{3}{2}(\gamma_{12})^3}{Q_2} + \frac{(\beta_{2\overline{2}})^\frac{3}{2} (\gamma_{12})^\frac{5}{2}}{Q_1}  \\ & + \frac{(\beta_{1\overline{2}})^\frac{3}{2} (\gamma_{12})^\frac{3}{2}}{Q_2}  - \frac{\overline{Q}_{2}}{{Q_2}^2} (\gamma_{12})^3  \left(\frac{\beta_{1\overline{2}}}{\gamma_{12}}\right) - \frac{(\beta_{1\overline{2}})^\frac{3}{2} \gamma_{12}}{Q_1}  + \frac{(\gamma_{12})^\frac{5}{2} (\beta_{2\overline{2}})^\frac{3}{2}}{Q_1} \\ & + \frac{\overline{Q}_{2}}{{Q_1}^2} (\gamma_{12})^2  \left(\frac{\beta_{1\overline{2}}}{\gamma_{12}}\right)  - \frac{Q_2}{{Q_1}^2} (\beta_{2\overline{2}})^\frac{3}{2} (\gamma_{12})^2 \Biggr)   + \frac{N_2}{3 \eta} \Biggl( \frac{(\gamma_{12})^3}{Q_2}  - \frac{2(\gamma_{12})^\frac{5}{2}}{Q_1} \\ & + \frac{Q_2}{{Q_1}^2} (\gamma_{12})^2 \Biggr)   + \frac{5}{21 \eta} \Biggl(\frac{N_2}{Q_2} (\gamma_{12})^3 + \frac{\overline{N}_{2} \overline{Q}_{2}}{{Q_2}^2}(\gamma_{12})^3  \left(\frac{\beta_{1\overline{2}}}{\gamma_{12}}\right) \\ & - \frac{\overline{N}_{2}}{Q_2} (\beta_{2\overline{2}})^\frac{3}{2} (\gamma_{12})^3    - \frac{N_2 Q_2}{{Q_1}^2} (\gamma_{12})^2  - \frac{\overline{N}_{2} \overline{Q}_{2}}{{Q_1}^2} (\gamma_{12})^2  \left(\frac{\beta_{1\overline{2}}}{\gamma_{12}}\right) \\ & + \frac{\overline{N}_{2} Q_2}{{Q_1}^2} (\gamma_{12})^2  \left. (\beta_{2\overline{2}})^\frac{3}{2} \Biggr) \Biggr] \frac{df_2(x)}{dx} \right|_{x_2} + \Biggl[ \frac{N_1}{3 \eta Q_1} - \frac{N_2 Q_2}{3 \eta {Q_1}^2} \gamma_{12}   + \frac{N_{2}}{3\eta Q_{1}} (\gamma_{12})^\frac{3}{2} \\ & + \frac{\overline{N}_{2}}{3 \eta Q_1} \Biggl( (\beta_{1\overline{2}})^\frac{3}{2} - (\gamma_{12})^\frac{3}{2} (\beta_{2\overline{2}})^\frac{3}{2} - \frac{\overline{Q}_{2}}{Q_1}  \left(\frac{\beta_{1\overline{2}}}{\gamma_{12}}\right) \gamma_{12}               
             +  \frac{Q_2}{Q_1} (\beta_{2\overline{2}})^\frac{3}{2} \gamma_{12} \Biggr) \\ & + \frac{5}{21 \eta} \Biggl( \frac{N_1}{Q_1} + \frac{N_2 Q_2}{{Q_1}^2} \gamma_{12}  - \frac{N_2}{Q_1} (\gamma_{12})^\frac{3}{2}   + \frac{\overline{N}_{2}}{Q_1} (\gamma_{12})^\frac{3}{2} (\beta_{2\overline{2}})^\frac{3}{2}  \\ & - \frac{\overline{N}_{2}}{Q_1} (\beta_{1\overline{2}})^\frac{3}{2} -  \frac{\overline{N}_{2} Q_2}{{Q_1}^2} (\beta_{2\overline{2}})^\frac{3}{2} \gamma_{12}  + \frac{\overline{N}_{2} \overline{Q}_{2}}{{Q_1}^2} \gamma_{12}  \left(\frac{\beta_{1\overline{2}}}{\gamma_{12}}\right) \Biggr) \left. \Biggr] \frac{df_1(x)}{dx}\right|_{x_1},
             \label{eqn:k11_caseI}
            \end{split}
            \end{equation}
            
        \begin{equation}
            C = \frac{12\hbar^2\sqrt{R}}{20 \pi m_1}\biggl[\frac{2\times(4/3)\alpha_s}{a}\biggr]^\frac{5}{2}.
            \label{eqn:expression_of_C}
        \end{equation} 
            \begin{equation}
            \begin{split}     
            K_{\overline{2}\overline{2}} 
            &= - \left. \frac{4}{7\overline{Q}_{2}} \sqrt{x} (\overline{f}_{2}(x))^\frac{5}{2} \right|_{\overline{x}_{2}} + \frac{4 \overline{N}_{2}}{7 \eta \overline{Q}_{2}} \frac{d\overline{f}_{2}(x)}{dx}\biggr|_{\overline{x}_{2}}
              \end{split},              
            \label{eqn:k2bar2bar_caseI}
            \end{equation}
        \begin{equation}
            \begin{split}            
            K_{\overline{2}2} &= - \left. \frac{4}{7\overline{Q}_{2}}(\beta_{2\overline{2}})^\frac{3}{2}\sqrt{x}(\overline{f}_{2}(x))^\frac{5}{2}\right|_{\overline{x}_{2}}  +\frac{\overline{N}_{2}}{3\eta} \biggl[\frac{(\beta_{2\overline{2}})^\frac{3}{2}}{\overline{Q}_{2}}-\frac{\left(\frac{\beta_{1\overline{2}}}{\gamma_{12}}\right)}{Q_2} \\ 
            & +\left. \frac{5(\beta_{2\overline{2}})^\frac{3}{2}}{7\overline{Q}_{2}}\biggr]\frac{d\overline{f}_{2}(x)}{dx}\right|_{\overline{x}_{2}}  + \left. \frac{\overline{N}_{2}}{3\eta Q_2} \frac{df_2(x)}{dx}\right|_{x_2}.
            \label{eqn:k2bar2_caseI}
            \end{split}
        \end{equation}
            \begin{equation}
                \begin{split}
                    K_{\overline{2}1}  &= -  \left.\frac{4}{7\overline{Q}_{2}} (\beta_{1\overline{2}})^\frac{3}{2} \sqrt{x} (\overline{f}_{2}(x))^\frac{5}{2} \right|_{\overline{x}_{2}}  + \frac{\overline{N}_{2}}{3 \eta} \Biggl[ \frac{5}{7 \overline{Q}_{2}} (\beta_{1\overline{2}})^\frac{3}{2}+ \frac{(\beta_{1\overline{2}})^\frac{3}{2}}{\overline{Q}_{2}}  \\ & - \frac{(\gamma_{12})^\frac{3}{2} \left(\frac{\beta_{1\overline{2}}}{\gamma_{12}}\right)}{Q_{2}} \Biggr] \left. \frac{d\overline{f}_{2}(x)}{dx}\right|_{\overline{x}_{2}}   +   \frac{\overline{N}_{2}}{3 \eta}  \Biggl[\frac{(\gamma_{12})^\frac{3}{2}}{Q_2} - \left. \frac{\gamma_{12}}{Q_1} \Biggr] \frac{df_2(x)}{dx} \right|_{x_2} \\ & + \left. \frac{\overline{N}_{2}}{3 \eta Q_1} \frac{df_1(x)}{dx} \right|_{x_1},
                \end{split}
            \label{eqn:k2bar1_caseI}
            \end{equation} 
            \begin{equation}
                \begin{split}
                          K_{22} &= \frac{4}{7} \biggl( \frac{\left(\frac{\beta_{1\overline{2}}}{\gamma_{12}}\right)^\frac{5}{2}}{Q_{2}} - \frac{\left(\beta_{2\overline{2}
                          }\right)^{3}}{\overline{Q}_{2}} \left. \biggr) \sqrt{x}  (\overline{f}_{2}(x))^\frac{5}{2} \right|_{\overline{x}_{2}}  - \left. \frac{4}{7 Q_{2}} \sqrt{x} (f_{2}(x))^\frac{5}{2} \right|_{x_{2}} \\ & + \frac{\overline{N}_{2}}{3 \eta} \biggl[\left(\frac{\beta_{1\overline{2}}}{\gamma_{12}}\right)^{2} \frac{\overline{Q}_{2}}{Q_{2}^{2}}  - \frac{(\beta_{2\overline{2}})^\frac{3}{2} \left(\frac{\beta_{1\overline{2}}}{\gamma_{12}}\right)}{Q_{2}}   + \frac{5}{7 \overline{Q}_{2}} (\beta_{2\overline{2}})^{3}  \\ & - \frac{5 \overline{Q}_{2}}{7 Q_{2}^{2}} \left(\frac{\beta_{1\overline{2}}}{\gamma_{12}}\right)^{2} \left. \biggr] \frac{d\overline{f}_{2}(x)}{dx} \right|_{\overline{x}_{2}} + \biggl[ \frac{5 N_{2}}{21 \eta Q_{2}}  + \beta_{2\overline{2}} \frac{\overline{N}_{2}}{3 \eta Q_{2}} \biggl( (\beta_{2\overline{2}})^\frac{1}{2} \\ & - \frac{5}{7} (\beta_{2\overline{2}})^\frac{1}{2} \biggr)  - \frac{\overline{N}_{2} \overline{Q}_{2}}{3 \eta Q_{2}^{2}}   \left(\frac{\beta_{1\overline{2}}}{\gamma_{12}}\right)   + \frac{5 \overline{N}_{2} \overline{Q}_{2}}{21 \eta Q_{2}^{2}} \left(\frac{\beta_{1\overline{2}}}{\gamma_{12}}\right) \left. \biggr] \frac{df_{2}(x)}{dx} \right|_{x_{2}} \\ & + (\beta_{2\overline{2}})^3\frac{\overline{N}_{2}}{3 \eta \overline{Q}_{2} } \left. \frac{d\overline{f}_{2}(x)}{dx} \right|_{\overline{x}_{2}} - \left. \left(\frac{\beta_{1\overline{2}}}{\gamma_{12}}\right) (\beta_{2\overline{2}})^\frac{3}{2}\frac{\overline{N}_{2}}{3 \eta Q_2} \frac{d \overline{f}_{2}(x)}{dx} \right|_{\overline{x}_{2}} \\ & + \left. \frac{1}{Q_2} \frac{N_2}{3\eta} \frac{df_2(x)}{dx} \right|_{x_2},
                \end{split}
            \label{eqn:k22_caseI}
            \end{equation}

\begin{equation}
\begin{split}
K_{12} &= K_{21} =
\left.\frac{4}{7 Q_2}
\left(\frac{\beta_{1\overline{2}}}{\gamma_{12}}\right)^{\frac{5}{2}}
\gamma_{12}^{\frac{3}{2}} \sqrt{x}\,\overline{f}_{2}(x)^{\frac{5}{2}}
\right|_{\overline{x}_{2}} 
- \left.
\frac{4}{7 \overline{Q}_{2}}
\beta_{1\overline{2}}^{\frac{3}{2}}
\beta_{2\overline{2}}^{\frac{3}{2}}
\sqrt{x}\,\overline{f}_{2}(x)^{\frac{5}{2}}
\right|_{\overline{x}_{2}} 
\\ &- \left.
\frac{4}{7 Q_2}\,
\gamma_{12}^{\frac{3}{2}}
\sqrt{x}\,f_2(x)^{\frac{5}{2}}
\right|_{x_2} 
+ \left.
\frac{5 \overline{N}_{2}}{21 \eta \overline{Q}_{2}}
\beta_{1\overline{2}}^{\frac{3}{2}}
\beta_{2\overline{2}}^{\frac{3}{2}}
\frac{d\overline{f}_{2}(x)}{dx}
\right|_{\overline{x}_{2}} 
\\ & - \left.
\frac{\overline{N}_{2}}{3 \eta Q_2}
\gamma_{12}^{\frac{3}{2}}
\left(\frac{\beta_{1\overline{2}}}{\gamma_{12}}\right)
 \beta_{2\overline{2}}^{\frac{3}{2}}
\frac{d\overline{f}_{2}(x)}{dx}
\right|_{\overline{x}_{2}} 
- \left.
\frac{5 \overline{N}_{2} \overline{Q}_{2}}{21 \eta Q_2^{2}}
\left(\frac{\beta_{1\overline{2}}}{\gamma_{12}}\right)^{2}
\gamma_{12}^{\frac{3}{2}}
\frac{d\overline{f}_{2}(x)}{dx}
\right|_{\overline{x}_{2}} 
\\ & + \left.
\frac{\overline{N}_{2} \overline{Q}_{2}}{3 \eta Q_2^{2}}
\left(\frac{\beta_{1\overline{2}}}{\gamma_{12}}\right)^{2}
\gamma_{12}^{\frac{3}{2}}
\frac{d\overline{f}_{2}(x)}{dx}
\right|_{\overline{x}_{2}} 
 + \left.
\frac{\overline{N}_{2}}{3 \eta \overline{Q}_{2}}
\beta_{1\overline{2}}^{\frac{3}{2}}
\beta_{2\overline{2}}^{\frac{3}{2}}
\frac{d\overline{f}_{2}(x)}{dx}
\right|_{\overline{x}_{2}} 
\\ & - \left.
\frac{\overline{N}_{2}}{3 \eta Q_2}
\beta_{1\overline{2}}^{\frac{3}{2}}
\left(\frac{\beta_{1\overline{2}}}{\gamma_{12}}\right)
\frac{d\overline{f}_{2}(x)}{dx}
\right|_{\overline{x}_{2}} 
+ \left.
\frac{5 N_2}{21 \eta Q_2}
\gamma_{12}^{\frac{3}{2}}
\frac{d f_2(x)}{dx}
\right|_{x_2} 
\\ & + \left.
\frac{5 \overline{N}_{2} \overline{Q}_{2}}{21 \eta Q_2^{2}}
\gamma_{12}^{\frac{3}{2}}
\left(\frac{\beta_{1\overline{2}}}{\gamma_{12}}\right)
\frac{d f_2(x)}{dx}
\right|_{x_2} 
- \left.
\frac{5 \overline{N}_{2}}{21 \eta Q_2}
\beta_{2\overline{2}}^{\frac{3}{2}}
\gamma_{12}^{\frac{3}{2}}
\frac{d f_2(x)}{dx}
\right|_{x_2} 
\\ & - \left.
\frac{\overline{N}_{2} \overline{Q}_{2}}{3 \eta Q_2^{2}}
\gamma_{12}^{\frac{3}{2}}
\left(\frac{\beta_{1\overline{2}}}{\gamma_{12}}\right)
\frac{d f_2(x)}{dx}
\right|_{x_2} 
- \left.
\frac{N_2}{3 \eta Q_1}
\gamma_{12}
\frac{d f_2(x)}{dx}
\right|_{x_2} 
\\
&+ \left.
\frac{\overline{N}_{2}}{3 \eta Q_2}
\beta_{1\overline{2}}^{\frac{3}{2}}
\frac{d f_2(x)}{dx}
\right|_{x_2} 
+ \left.
\frac{N_2}{3 \eta Q_2}
\gamma_{12}^{\frac{3}{2}}
\frac{d f_2(x)}{dx}
\right|_{x_2} 
+ \left.
\frac{N_2}{3 \eta Q_1}
\frac{d f_1(x)}{dx}
\right|_{x_1}.
\end{split}
\label{eqn:k12_caseI}
\end{equation}

\section{} \label{appendix II}
    The expression of kinetic energy for Case II is 
    \begin{equation}
            \begin{split}
                K.E_{case\text{ }II} &= C \biggl[\frac{4}{7} \biggl[ \alpha_{1}g_{1} \biggl( (\beta_{12})^\frac{5}{2} - (\gamma_{1\overline{2}})^\frac{5}{2} \left( \frac{\beta_{12}}{\gamma_{1\overline{2}}}\right)^\frac{5}{2} \biggr)  + \alpha_{2}g_{2} \biggr] \\ & \left. \sqrt{x} (f_{2}(x))^\frac{5}{2} \right|_{x_{2}}  + \left. \frac{4}{7} \overline{\alpha}_{2} \overline{g}_{2} \sqrt{x} (\overline{f}_{2}(x))^\frac{5}{2} \right|_{\overline{x}_{2}}  + \left. \frac{4}{7} \alpha_{1} g_{1} \sqrt{x} (f_{1}(x))^\frac{5}{2} \right|_{x_{1}} \\ & + \frac{5 N_{2}}{21 \eta} \biggl[ \left( \frac{\beta_{12}}{\gamma_{1\overline{2}}}\right)^{2} \frac{Q_{2}}{\overline{Q}_{2}} \biggl(\alpha_{1} g_{1} (\gamma_{1\overline{2}})^\frac{5}{2}  + \overline{\alpha}_{2} \overline{g}_{2} \biggr) -\alpha_{1} g_{1} (\beta_{12})^\frac{5}{2} \\ & - \overline{\alpha}_{2} \overline{g}_{2} (\beta_{\overline{2}2})^\frac{5}{2} - \alpha_{2} g_{2} \biggr] \left. \frac{df_{2}(x)}{dx} \right|_{x_{2}} + \frac{5}{21 \eta}  \biggl[ \frac{(\gamma_{1\overline{2}})^{2}}{Q_{1}} \alpha_{1}g_{1}   \biggl[ \overline{N}_{2} \overline{Q}_{2} \\ & + \left( \frac{\beta_{12}}{\gamma_{1\overline{2}}}\right) N_{2}Q_{2} - (\beta_{\overline{2}2})^{\frac{3}{2}}  N_{2}\overline{Q}_{2} \biggr]  - \biggl( \alpha_{1} g_{1} (\gamma_{1\overline{2}})^\frac{5}{2} \\ & + \overline{\alpha}_{2} \overline{g}_{2} \biggr)  \biggl[\overline{N}_{2} + \left( \frac{\beta_{12}}{\gamma_{1\overline{2}}}\right) \frac{N_{2} Q_{2}}{\overline{Q}_{2}}  - (\beta_{\overline{2}2})^\frac{3}{2} N_{2} \biggr] \left. \biggr] \frac{d\overline{f}_{2}(x)}{dx} \right|_{\overline{x}_{2}} \\ &  - \frac{5 \alpha_{1}g_{1}}{21 \eta} \biggl[ N_{1} + \gamma_{1\overline{2}}  \frac{\overline{N}_{2} \overline{Q}_{2}}{Q_{1}} + (\gamma_{1\overline{2}})^{\frac{3}{2}} \biggl( (\beta_{\overline{2}2})^\frac{3}{2} N_{2} \\ & -  \overline{N}_{2} \biggr)  - (\beta_{12})^\frac{3}{2} N_{2} - (\gamma_{1\overline{2}})(\beta_{\overline{2}2})^\frac{3}{2} \frac{N_{2} \overline{Q}_{2}}{Q_{1}} \\ &  + \gamma_{1\overline{2}} \left( \frac{\beta_{12}}{\gamma_{1\overline{2}}}\right) \frac{N_{2} Q_{2}}{Q_{1}} \left. \biggr] \frac{df_{1}(x)}{dx} \right|_{x_{1}} \biggr],
            \end{split}
        \label{eqn:kinetic_energy_caseII}
        \end{equation}
             The final expression of constants that appear in the potential energy term for Case II are shown below.    
            \begin{equation*}
                \begin{split}
K_{11} & = \frac{4}{7} \biggl[ \frac{\left( \frac{\beta_{12}}{\gamma_{1\overline{2}}}\right)^\frac{5}{2} (\gamma_{1\overline{2}})^3}{\overline{Q}_{2}} - \left. \frac{(\beta_{12})^3}{Q_2} \biggr] \sqrt{x} (f_2(x))^\frac{5}{2} \right|_{x_2}  +  \frac{4}{7} \biggl[\frac{(\gamma_{1\overline{2}})^\frac{5}{2}}{Q_1} \\ & - \left. \frac{(\gamma_{1\overline{2}})^3}{\overline{Q}_{2}} \biggr] \sqrt{x} (\overline{f}_{2}(x))^\frac{5}{2} \right|_{\overline{x}_{2}}  -\left. \frac{4}{7 Q_1} \sqrt{x} (f_1(x))^\frac{5}{2} \right|_{x_1}  \\ & + \Biggl[ \frac{N_2}{3 \eta} \Biggl( \frac{(\beta_{12})^3}{Q_2}  - \frac{(\beta_{12})^\frac{3}{2} (\gamma_{1\overline{2}})^\frac{3}{2} \left( \frac{\beta_{12}}{\gamma_{1\overline{2}}}\right)}{\overline{Q}_{2}} \\ &+ \frac{Q_2}{{\overline{Q}_{2}}^2} (\gamma_{1\overline{2}})^3 \left( \frac{\beta_{12}}{\gamma_{1\overline{2}}}\right)^2 - \frac{(\beta_{12})^\frac{3}{2} (\gamma_{1\overline{2}})^\frac{3}{2} \left( \frac{\beta_{12}}{\gamma_{1\overline{2}}}\right)}{\overline{Q}_{2}} \Biggr) \\ & + \frac{5 N_2}{21 \eta} \Biggl( \frac{(\beta_{12})^3}{Q_2}   - \left. \frac{Q_2}{{\overline{Q}_{2}}^2} \left( \frac{\beta_{12}}{\gamma_{1\overline{2}}}\right)^2 (\gamma_{1\overline{2}})^3 \Biggr) \Biggr] \frac{df_2(x)}{dx} \right|_{x_2}  \\ & + \Biggl[ \frac{N_2}{3 \eta} \Biggl( \frac{(\beta_{12})^\frac{3}{2} (\gamma_{1\overline{2}})^\frac{3}{2}}{\overline{Q}_{2}} - \frac{(\beta_{12})^\frac{3}{2} \gamma_{1\overline{2}}}{Q_1} - \frac{(\beta_{\overline{2}2})^\frac{3}{2}(\gamma_{1\overline{2}})^3}{\overline{Q}_{2}} \\ & + \frac{(\beta_{\overline{2}2})^\frac{3}{2} (\gamma_{1\overline{2}})^\frac{5}{2}}{Q_1}   + \frac{(\beta_{12})^\frac{3}{2} (\gamma_{1\overline{2}})^\frac{3}{2}}{\overline{Q}_{2}} - \frac{Q_2}{{\overline{Q}_{2}}^2} (\gamma_{1\overline{2}})^3 \left( \frac{\beta_{12}}{\gamma_{1\overline{2}}}\right) \\ & - \frac{(\beta_{12})^\frac{3}{2} \gamma_{1\overline{2}}}{Q_1}  + \frac{(\gamma_{1\overline{2}})^\frac{5}{2} (\beta_{\overline{2}2})^\frac{3}{2}}{Q_1}  + \frac{Q_2}{{Q_1}^2} (\gamma_{1\overline{2}})^2 \left( \frac{\beta_{12}}{\gamma_{1\overline{2}}}\right) \\ & - \frac{\overline{Q}_{2}}{{Q_1}^2} (\beta_{\overline{2}2})^\frac{3}{2} (\gamma_{1\overline{2}})^2 \Biggr)   + \frac{\overline{N}_{2}}{3 \eta} \Biggl( \frac{(\gamma_{1\overline{2}})^3}{\overline{Q}_{2}} - \frac{2(\gamma_{1\overline{2}})^\frac{5}{2}}{Q_1} \\ & + \frac{\overline{Q}_{2}}{{Q_1}^2} (\gamma_{1\overline{2}})^2 \Biggr)  + \frac{5}{21 \eta} \Biggl( \frac{\overline{N}_{2}}{\overline{Q}_{2}} (\gamma_{1\overline{2}})^3 + \frac{N_2 Q_2}{{\overline{Q}_{2}}^2}(\gamma_{1\overline{2}})^3 \left( \frac{\beta_{12}}{\gamma_{1\overline{2}}}\right) \\ &- \frac{N_2}{\overline{Q}_{2}} (\beta_{\overline{2}2})^\frac{3}{2} (\gamma_{1\overline{2}})^3 - \frac{\overline{N}_{2} \overline{Q}_{2}}{{Q_1}^2} (\gamma_{1\overline{2}})^2  - \frac{N_2 Q_2}{{Q_1}^2} (\gamma_{1\overline{2}})^2  \\ & \times \left( \frac{\beta_{12}}{\gamma_{1\overline{2}}}\right)   + \frac{N_2 \overline{Q}_{2}}{{Q_1}^2} (\gamma_{1\overline{2}})^2  (\beta_{\overline{2}2})^\frac{3}{2} \Biggr) \Biggr] \left. \frac{d\overline{f}_{2}(x)}{dx} \right|_{\overline{x}_{2}} + \Biggl[ \frac{N_1}{3 \eta Q_1} \\ & - \frac{\overline{N}_{2} \overline{Q}_{2}}{3 \eta {Q_1}^2} \gamma_{1\overline{2}}  + \frac{\overline{N}_{2}}{3\eta Q_{1}} (\gamma_{1\overline{2}})^\frac{3}{2} + \frac{N_2}{3 \eta Q_1} \Biggl( (\beta_{12})^\frac{3}{2}
                    \end{split}
                    \end{equation*}      
                    
                    \begin{equation}
                    \begin{split}
                      &- (\gamma_{1\overline{2}})^\frac{3}{2} (\beta_{\overline{2}2})^\frac{3}{2} - \frac{Q_2}{Q_1} \left( \frac{\beta_{12}}{\gamma_{1\overline{2}}}\right) \gamma_{1\overline{2}}  + \frac{\overline{Q}_{2}}{Q_1} (\beta_{\overline{2}2})^\frac{3}{2} \gamma_{1\overline{2}} \Biggr) \\ & + \frac{5}{21 \eta} \Biggl( \frac{N_1}{Q_1} + \frac{\overline{N}_{2} \overline{Q}_{2}}{{Q_1}^2} \gamma_{1\overline{2}}  - \frac{\overline{N}_{2}}{Q_1} (\gamma_{1\overline{2}})^\frac{3}{2}   + \frac{N_2}{Q_1} (\gamma_{1\overline{2}})^\frac{3}{2} (\beta_{\overline{2}2})^\frac{3}{2} \\ & - \frac{N_2}{Q_1} (\beta_{12})^\frac{3}{2}  - \left. \frac{N_2 \overline{Q}_{2}}{{Q_1}^2} (\beta_{\overline{2}2})^\frac{3}{2} \gamma_{1\overline{2}}  + \frac{N_2 Q_2}{{Q_1}^2} \gamma_{1\overline{2}}  \left( \frac{\beta_{12}}{\gamma_{1\overline{2}}}\right) \Biggr) \Biggr] \frac{df_1(x)}{dx}\right|_{x_1},
                \end{split}
              \label{eqn:k11_caseII}
            \end{equation} 
            \begin{equation}
            K_{22} = - \left. \frac{4}{7Q_{2}} \sqrt{x} (f_{2}(x))^\frac{5}{2} \right|_{x_{2}} + \frac{4 N_{2}}{7 \eta Q_{2}} \frac{df_{2}(x)}{dx}\biggr|_{x_{2}},
            \label{eqn:k22_caseII}
            \end{equation}
            \begin{equation}
                \begin{split}
                    K_{\overline{2}\overline{2}} &= \frac{4}{7} \biggl( \frac{\left(\frac{\beta_{12}}{\gamma_{1\overline{2}}}\right)^\frac{5}{2}}{\overline{Q}_{2}} - \frac{\left(\beta_{\overline{2}2
                    }\right)^{3}}{Q_{2}} \left. \biggr) \sqrt{x}  (f_{2}(x))^\frac{5}{2} \right|_{x_{2}} - \left. \frac{4}{7 \overline{Q}_{2}} \sqrt{x} (\overline{f}_{2}(x))^\frac{5}{2} \right|_{\overline{x}_{2}} \\ & + \frac{N_{2}}{3 \eta} \biggl[ \left(\frac{\beta_{12}}{\gamma_{1\overline{2}}}\right)^{2} \frac{Q_{2}}{\overline{Q}_{2}^{2}} - \frac{(\beta_{\overline{2}2})^\frac{3}{2} \left(\frac{\beta_{12}}{\gamma_{1\overline{2}}}\right)}{\overline{Q}_{2}}  + \frac{5}{7 Q_{2}} (\beta_{\overline{2}2})^{3} \\ & - \frac{5 Q_{2}}{7 \overline{Q}_{2}^{2}} \left(\frac{\beta_{12}}{\gamma_{1\overline{2}}}\right)^{2} \left. \biggr] \frac{df_{2}(x)}{dx} \right|_{x_{2}} + \biggl[ \frac{5 \overline{N}_{2}}{21 \eta \overline{Q}_{2}}  + \beta_{\overline{2}2} \frac{N_{2}}{3 \eta \overline{Q}_{2}} \\ & \times \biggl( (\beta_{\overline{2}2})^\frac{1}{2} - \frac{5}{7} (\beta_{\overline{2}2})^\frac{1}{2} \biggr) - \frac{N_{2} Q_{2}}{3 \eta \overline{Q}_{2}^{2}}  \left(\frac{\beta_{12}}{\gamma_{1\overline{2}}}\right) \\ & + \frac{5 N_{2} Q_{2}}{21 \eta \overline{Q}_{2}^{2}} \left(\frac{\beta_{12}}{\gamma_{1\overline{2}}}\right) \left. \biggr] \frac{d\overline{f}_{2}(x)}{dx} \right|_{\overline{x}_{2}}  + (\beta_{\overline{2}2})^3\frac{N_{2}}{3 \eta Q_{2} } \left. \frac{df_{2}(x)}{dx} \right|_{x_{2}} \\ & -  \left(\frac{\beta_{12}}{\gamma_{1\overline{2}}}\right) (\beta_{\overline{2}2})^\frac{3}{2} \frac{N_{2}}{3 \eta \overline{Q}_{2}} \left. \frac{d f_{2}(x)}{dx} \right|_{x_{2}} + \left. \frac{\overline{N}_{2}}{3\eta \overline{Q}_{2}} \frac{d\overline{f}_{2}(x)}{dx} \right|_{\overline{x}_{2}},
                \end{split}
            \label{eqn:k2bar2bar_caseII}
            \end{equation}
            \begin{equation}
                \begin{split}
                    K_{21} & = K_{12}= -  \left.\frac{4}{7Q_{2}} (\beta_{12})^\frac{3}{2} \sqrt{x} (f_{2}(x))^\frac{5}{2} \right|_{x_{2}} + \frac{N_{2}}{3 \eta} \Biggl[ \frac{5}{7 Q_{2}} (\beta_{12})^\frac{3}{2} \\ & +  \frac{(\beta_{12})^\frac{3}{2}}{Q_{2}}  - \frac{(\gamma_{1\overline{2}})^\frac{3}{2} \left(\frac{\beta_{12}}{\gamma_{1\overline{2}}}\right)}{\overline{Q}_{2}} \Biggr] \left. \frac{df_{2}(x)}{dx}\right|_{x_{2}} \quad +   \frac{N_{2}}{3 \eta}  \Biggl(\frac{(\gamma_{1\overline{2}})^\frac{3}{2}}{\overline{Q}_{2}}  \\ & - \left. \frac{\gamma_{1\overline{2}}}{Q_1} \Biggr) \frac{d\overline{f}_{2}(x)}{dx} \right|_{\overline{x}_{2}}  + \left. \frac{N_{2}}{3 \eta Q_1} \frac{df_1(x)}{dx} \right|_{x_1},
                \end{split}
            \label{eqn:k21_caseII}
            \end{equation}
            \begin{equation}
                \begin{split}
                    K_{\overline{2}1}&= \left.\frac{4}{7\overline{Q}_{2}} \left(\frac{\beta_{12}}{\gamma_{1\overline{2}}}\right)^\frac{5}{2} (\gamma_{1\overline{2}})^\frac{3}{2} \sqrt{x} (f_{2}(x))^\frac{5}{2} \right|_{x_{2}} - \frac{4}{7 Q_{2}} (\beta_{12})^\frac{3}{2} (\beta_{\overline{2}2})^\frac{3}{2} \\ & \times \left. \sqrt{x} (f_{2}(x))^\frac{5}{2} \right|_{x_{2}}  - \frac{4}{7 \overline{Q}_{2}} (\gamma_{1\overline{2}})^\frac{3}{2} \sqrt{x} \left. (\overline{f}_{2}(x))^\frac{5}{2} \right|_{\overline{x}_{2}} \\ & +  \frac{5 N_{2}}{21 \eta Q_{2}} (\beta_{12})^\frac{3}{2}  (\beta_{\overline{2}2})^\frac{3}{2} \left. \frac{df_{2}(x)}{dx} \right|_{x_{2}} -  \frac{N_{2}}{3 \eta \overline{Q}_{2}} (\gamma_{1\overline{2}})^\frac{3}{2} \left(\frac{\beta_{12}}{\gamma_{1\overline{2}}}\right) \\ & \times (\beta_{\overline{2}2})^\frac{3}{2} \left. \frac{df_{2}(x)}{dx} \right|_{x_{2}} -  \frac{5 N_{2} Q_{2}}{21 \eta \overline{Q}_{2}^2 } \left(\frac{\beta_{12}}{\gamma_{1\overline{2}}}\right)^2  (\gamma_{1\overline{2}})^\frac{3}{2} \left. \frac{df_{2}(x)}{dx} \right|_{x_{2}} \\ & +  \frac{N_{2} Q_{2}}{3 \eta \overline{Q}_{2}^2} (\gamma_{1\overline{2}})^\frac{3}{2}  \left(\frac{\beta_{12}}{\gamma_{1\overline{2}}}\right)^2 \left. \frac{df_{2}(x)}{dx} \right|_{x_{2}} +  \frac{N_{2}}{3 \eta Q_{2}} (\beta_{12})^\frac{3}{2} (\beta_{\overline{2}2})^\frac{3}{2} \\ & \times \left. \frac{df_{2}(x)}{dx} \right|_{x_{2}} - \left. \frac{N_{2}}{3 \eta \overline{Q}_{2}} (\beta_{12})^\frac{3}{2} \left(\frac{\beta_{12}}{\gamma_{1\overline{2}}}\right) \frac{df_{2}(x)}{dx} \right|_{x_{2}}  \\ & + \left. \frac{5\overline{N}_{2}}{21 \eta \overline{Q}_{2}} (\gamma_{1\overline{2}})^\frac{3}{2} \frac{d\overline{f}_{2}(x)}{dx} \right|_{\overline{x}_{2}}  +   \frac{5 N_{2} Q_{2}}{21 \eta \overline{Q}_{2}^2} (\gamma_{1\overline{2}})^\frac{3}{2}  \left(\frac{\beta_{12}}{\gamma_{1\overline{2}}}\right) \left. \frac{d\overline{f}_{2}(x)}{dx} \right|_{\overline{x}_{2}} \\ & -  \frac{5 N_{2}}{21 \eta \overline{Q}_{2}} (\beta_{\overline{2}2})^\frac{3}{2}  (\gamma_{1\overline{2}})^\frac{3}{2} \left. \frac{d\overline{f}_{2}(x)}{dx}\right|_{\overline{x}_{2}}  -  \frac{N_{2}Q_{2}}{3 \eta \overline{Q}_{2}^2} (\gamma_{1\overline{2}})^\frac{3}{2} \left(\frac{\beta_{12}}{\gamma_{1\overline{2}}}\right) \\ & \times \left. \frac{d\overline{f}_{2}(x)}{dx} \right|_{\overline{x}_{2}} - \left. \frac{\overline{N}_{2}}{3 \eta Q_1} \gamma_{1\overline{2}} \frac{d\overline{f}_{2}(x)}{dx} \right|_{\overline{x}_{2}}   + \left. \frac{N_{2}}{3 \eta \overline{Q}_{2}} (\beta_{12})^\frac{3}{2} \frac{d\overline{f}_{2}(x)}{dx} \right|_{\overline{x}_{2}} \\ & +  \frac{\overline{N}_{2}}{3 \eta \overline{Q}_{2}} (\gamma_{1\overline{2}})^\frac{3}{2}  \left. \frac{d\overline{f}_{2}(x)}{dx}\right|_{\overline{x}_{2}} + \left. \frac{\overline{N}_{2}}{3 \eta Q_1} \frac{df_1(x)}{dx} \right|_{x_1},
                \end{split}
            \label{eqn:k2bar1_caseII}
           \end{equation}       
            \begin{equation}
            \begin{split}
            K_{\overline{2}2} &= - \left. \frac{4}{7Q_{2}}(\beta_{\overline{2}2})^\frac{3}{2}\sqrt{x}(f_{2}(x))^\frac{5}{2}\right|_{x_{2}}  +\frac{N_{2}}{3\eta} \biggl(\frac{(\beta_{\overline{2}2})^\frac{3}{2}}{Q_{2}} -\frac{\left(\frac{\beta_{12}}{\gamma_{1\overline{2}}}\right)}{\overline{Q}_{2}} \\ &   +\left. \frac{5(\beta_{\overline{2}2})^\frac{3}{2}}{7Q_{2}}\biggr)\frac{df_{2}(x)}{dx}\right|_{x_{2}}  + \left. \frac{N_{2}}{3\eta \overline{Q}_{2}} \frac{d\overline{f}_{2}(x)}{dx}\right|_{\overline{x}_{2}}.
            \label{eqn:k2bar2_caseII}
            \end{split}
            \end{equation} 
            \label{appl2}



  \bibliographystyle{elsarticle-num} 



\bibliography{bibliography}
            
\end{document}